\documentclass[12pt]{article}

\usepackage[letterpaper,top=2cm,bottom=2cm,left=2cm,right=2cm,marginparwidth=1.75cm]{geometry}
\usepackage{amsmath,amssymb}
\usepackage{graphicx}
\usepackage[colorlinks=true, allcolors=blue]{hyperref}
\usepackage{authblk}
\usepackage{tabularray}
\usepackage{bbm}
\usepackage{multirow}
\usepackage{cite}
\usepackage{algorithm}
\usepackage{algpseudocode}
\UseTblrLibrary{booktabs}

\def\mathswitchr#1{\relax\ifmmode{\mathrm{#1}}\else$\mathrm{#1}$\fi}
\def\mathswitch#1{\relax\ifmmode#1\else$#1$\fi}
\newcommand{\brc}[1]{\left(#1\right)}

\newcommand{\corr}{\mathswitchr{Corr}}
\newcommand{\var}{\mathswitchr{var}}

\newcommand{\bgam}{\mathswitch{\boldsymbol{\gamma}}}

\newcommand{\bI}{\mathswitch{\boldsymbol{\mathrm{I}}}}

\newcommand{\bU}{\mathswitch{\boldsymbol{U}}}

\newcommand{\bt}{\mathswitch{\boldsymbol{t}}}
\newcommand{\bm}{\mathswitch{\boldsymbol{m}}}
\newcommand{\bn}{\mathswitch{\boldsymbol{n}}}

\newcommand{\declare}[2]{\vspace{2em}\noindent{\fontsize{14}{14}\selectfont\textbf{#1}}{%
\par\vspace{3pt}{\fontsize{12}{14}\selectfont #2}\par}}

\providecommand{\keywords}[1]
{
  \small
  \noindent
  \textbf{Keywords:} #1
}

\title{\textbf{\Large Confidence Intervals for the Risk Difference in Combined Unilateral and Bilateral Data Incorporating a Distribution-Based Approach}}
\author{Jia Zhou \footnote{\href{mailto:jiazhou@buffalo.edu}{jiazhou@buffalo.edu}} }
\author{Chang-Xing Ma \footnote{\href{mailto:cxma@buffalo.edu}{cxma@buffalo.edu}}}
\affil{Department of Biostatistics, University at Buffalo, Buffalo, NY 14214, USA}
\date{}

\begin{document}
\maketitle

\begin{abstract}
\fontsize{12pt}{14pt}\selectfont

Combined unilateral and bilateral binary outcomes frequently arise in studies involving paired organs. The risk difference is a clinically interpretable measure for comparing treatment effects between groups. Existing confidence interval methods are primarily based on asymptotic normality and may fail to adequately reflect finite-sample distributional features, particularly skewness.
To address this issue, we propose a distribution-based confidence interval derived from the probability distribution of the risk difference estimator and a modified MOVER procedure that accounts for intra-subject correlation. Their performances are compared with those of commonly used asymptotic methods through extensive simulation studies.
Across a broad range of parameter settings, all methods exhibited satisfactory performance as sample size increased. The proposed distribution-based interval achieved coverage probabilities close to the nominal level with interval widths comparable to those of existing procedures. In small sample settings, it was able to capture skewness in the sampling distribution that was not reflected by methods relying on asymptotic normality. Analyses of two real-world datasets demonstrated the practical applicability of the competing methods and yielded consistent inferential conclusions. The proposed approach provides an alternative framework for interval estimation of the risk difference in studies involving combined unilateral and bilateral binary outcomes.

\vspace{5mm}
\end{abstract}

\section{Introduction}
\label{sec:intro}
In comparative clinical studies involving paired organs, such as eyes or ears, binary outcomes are frequently correlated within subjects. Moreover, practical datasets often consist of a mixture of unilateral and bilateral observations, arising from circumstances such as missing measurements, prior surgical removal, and congenital absence of one organ. Properly analyzing such combined unilateral and bilateral data is essential for valid statistical inference, as discarding unilateral observations may lead to substantial loss of information and potential bias. 

A key challenge in analyzing correlated binary outcomes is appropriately accounting for intra-subject dependence. Several models have been proposed for this purpose. Rosner's constant $R$ model~\cite{Rosner_1982} and Dallal's conditional probability model~\cite{Dallal_1988} were among the earliest approaches, while Donner's constant correlation model~\cite{Donner1989rhoModel} parameterizes the within-subject dependence through a single correlation coefficient and has been shown to perform robustly across a wide range of scenarios~\cite{Thompson_1993}. More recently, the Clayton copula has been proposed to provide greater flexibility in modeling dependence between paired binary outcomes~\cite{liang2025testing,zhou2025goodness}. 

While much of the existing literature has focused on testing equality of proportions~\cite{Tang_2008,Pei_2008,ma2017rho,ma2022testing} or relative measures such as odds ratios~\cite{liu2013revisit,wang2016exactci,li2022statistical} and relative risks~\cite{klingenberg2010simultaneous,tian2025confidence,wang2025interval}, the risk difference remains a particularly appealing estimand in clinical and public health research~\cite{noordzij2017relative,laupacis1988assessment,jaeschke1995basic,schechtman2002odds,sinclair1994clinically}. As an absolute measure of treatment effect, the risk difference is directly interpretable and forms the basis of important quantities such as the number needed to treat~\cite{cook1995number,grieve2003number,mcalister2008number}. Consequently, reliable inference for the risk difference is of substantial practical importance.

In recent related work, Zhou and Ma~\cite{zhou2025testing} developed three likelihood-based asymptotic test statistics (the Wald-type, likelihood ratio, and score tests) for testing hypotheses concerning the risk difference between two proportions under Donner's model for combined unilateral and bilateral data. Extensive simulation studies demonstrated that all three tests maintain satisfactory control of the type I error, with the score test exhibiting slightly improved stability. Despite these encouraging results, hypothesis testing alone provides limited information in many applied settings, where interval estimation of the risk difference is often of primary interest.
The construction of confidence intervals for the risk difference in correlated binary data presents additional challenges. Standard asymptotic confidence intervals, typically derived from Wald, likelihood ratio, or score-based approaches, rely on large-sample normal approximations that may perform poorly in finite samples. Moreover, the sampling distribution of the risk difference under correlation is generally non-normal and can exhibit skewness or boundary effects, motivating the need for alternative inferential strategies.

In this paper, we develop a distribution-based approach for constructing confidence intervals for the risk difference. The proposed method is based on deriving the characteristic function of the risk difference and recovering its probability density function via the inverse Fourier transform. This framework enables direct approximation of the sampling distribution of the estimator without reliance on normal approximations, thereby facilitating confidence interval construction through distributional quantiles. The methodology applies to any dependence model for which the joint outcome probabilities can be specified. For concreteness and to facilitate comparison with existing asymptotic methods, we adopt Donner’s constant correlation model~\cite{Donner1989rhoModel} as the working dependence structure in this study, and also consider confidence intervals derived from the three commonly used asymptotic methods, i.e., the Wald-type, likelihood ratio, and score-based approaches, as well as that constructed by the method of variance estimates recovery (MOVER) procedure~\cite{Zou_2008con}.
Through simulation studies, we assess the performance of the proposed confidence intervals in terms of coverage probability and interval length, and compare them with their asymptotic counterparts across a wide range of parameter settings. Real-world examples from otolaryngologic and ophthalmologic studies are provided to illustrate the practical utility of the proposed approach.

The remainder of the paper is organized as follows. Section~\ref{sec:methods} introduces all the competing methods for confidence interval constructions, including the distribution-based approach, three asymptotic methods and adjusted MOVER procedures. Section~\ref{sec:results} consists of extensive simulation studies and real-world applications. Concluding remarks are given in Section~\ref{sec:conclusions}.

\section{Methods}
\label{sec:methods}
The risk difference concerns the difference in proportions of cured (or affected) organs (denoted by $\pi_i$) between the treatment ($i=2$) and control ($i=1$) groups, i.e., $\delta=\pi_2-\pi_1$. Let $Z_{ijk_j}$ denote the binary response ($1$ for the occurrence of a particular condition, and $0$ otherwise) of the $k_j$-th paired organ of the $j$-th subject in the $i$-th ($i=1,2$) group. Note that for subjects who contribute unilateral outcomes, $k_j$ collapses to either side of the paired organ and becomes a dummy index. Then $\pi_i$ is the marginal probability of the occurrence of the condition of interest (e.g., cured or affected), i.e., $Pr\brc{Z_{ijk_j}=1}=\pi_i$. Furthermore, let $\brc{m_{0i},m_{1i},m_{2i}}$ denote the bilateral counts in the $i$-th group, where $m_{ri}$ subjects have $r~(=0,1,2)$ organs cured (or affected). Similarly, let $\brc{n_{0i},n_{1i}}$ denote the unilateral counts in the $i$-th group. A $5\times2$ contingency table summarizing the combined bilateral and unilateral data structure is given in Table~\ref{tab:data_struc}, where the subscript `+' in the marginal totals indicates a sum of the corresponding index. 
\begin{table}[thpb]
    \centering
    \caption{Frequency table for the numbers of cured (or affected) organs for subjects in the control and treatment groups.}
    \label{tab:data_struc}
    \begin{tabular}{cccc}
    \toprule
    \# of cured or affected organs &control ($i=1$) &treatment ($i=2$) &total \\
    \midrule
    0 &$m_{01}$ &$m_{02}$ &$m_{0+}$ \\
    1 &$m_{11}$ &$m_{12}$ &$m_{1+}$ \\
    2 &$m_{21}$ &$m_{22}$ &$m_{2+}$ \\
    total &$m_{+1}$ &$m_{+2}$ &$m_{++}$ \\
    \midrule
    0 &$n_{01}$ &$n_{02}$ &$n_{0+}$ \\
    1 &$n_{11}$ &$n_{12}$ &$n_{1+}$ \\
    total &$n_{+1}$ &$n_{+2}$ &$n_{++}$ \\
    \bottomrule
    \end{tabular}
\end{table}

Conditioning on $m_{+i}$ and $n_{+i}$ in each group, the bilateral and unilateral counts respectively follow trinomial and binomial distributions, i.e.,
\begin{equation}
  \brc{m_{0i},m_{1i},m_{2i}}\sim Trinomial\brc{m_{+i},p_{0i},p_{1i},p_{2i}}, \quad
  n_{1i} \sim Binomial\brc{n_{+i},\pi_i}, 
  \label{eq:distribution}
\end{equation}
where the trinomial cell probabilities satisfy $\sum_{r=0}^2p_{ri}=1$ and can be expressed in terms of the marginal probability ($\pi_i$) and the intra-subject correlation ($\corr\brc{Z_{ijk_j},Z_{ij,3-k_j}}\equiv\rho_i$) as follows
\begin{equation}
  \begin{aligned}
    &p_{2i}=\pi_i\brc{\pi_i+\brc{1-\pi_i}\rho_i}, \\
    &p_{1i}=2\pi_i\brc{1-\pi_i}\brc{1-\rho_i}, \\
    &p_{0i}=\brc{1-\pi_i}\brc{1-\pi_i+\pi_i\rho_i}, 
  \end{aligned}
  \label{eq:prob_ri}
\end{equation}
where the intra-subject correlation is parameterized as $\rho_i=\rho$ under Donner's constant $\rho$ model~\cite{Donner1989rhoModel}, and of course could take different forms under other statistical models.

Based on the explicit distributions given above, quantities such as distribution and likelihood functions can be constructed accordingly. In what follows, we elaborate the distribution-based approach in Section~\ref{sec:methods:dist}, which uses the characteristic function and the inversion formula to reconstruct the sampling distribution of the risk difference estimator. In Section~\ref{sec:methods:asymp}, we review the commonly used asymptotic methods and the ways of constructing corresponding confidence intervals. The method of variance estimates recovery (MOVER) for constructing the confidence interval of the risk difference is discussed in Section~\ref{sec:methods:mover}.

\subsection{Distribution-based Approach}
\label{sec:methods:dist}
Let the risk difference estimator be $\tilde{\delta}=\tilde{\pi}_2-\tilde{\pi}_1$, where $\tilde{\pi}_i$ is the estimator of $\pi_i$, and takes the form
\begin{equation}
  \tilde{\pi}_i=\frac{m_{1i}+2m_{2i}+n_{1i}}{2m_{+i}+n_{+i}}, \quad i=1,2.
  \label{eq:pis}
\end{equation}

Then, conditioning on the total counts ($m_{+i},~n_{+i}$) in each group, $\tilde{\delta}$ is an unbiased estimator, i.e., $\mathbb{E}\brc{\tilde{\delta}}=\delta$.
The characteristic function of $\tilde{\delta}$ is given by
\begin{equation}
  \varphi_{\tilde{\delta}}\brc{t}=\mathbb{E}\brc{e^{i\tilde{\delta}t}}=\prod_{k=1}^2\brc{p_{0k}+p_{1k}~e^{s_k\frac{it}{D_k}}+p_{2k}~e^{s_k\frac{i2t}{D_k}}}^{m_{+k}}\brc{1-\pi_k+\pi_k~e^{s_k\frac{it}{D_k}}}^{n_{+k}},
  \label{eq:cf}
\end{equation}
where $s_k=\brc{-1}^k$ and $D_k=2m_{+k}+n_{+k}$. The derivation of the characteristic function can be found in Appendix~\ref{app:CF}. More specifically, under the reparameterization $\pi_2=\pi_1+\delta$, we have $\varphi_{\tilde{\delta}}\brc{t}=\varphi_{\tilde{\delta}}\brc{t;\delta,\pi_1,\rho}$.

Approximating the distribution of $\tilde{\delta}$ as continuous, the corresponding probability density function (PDF) can be obtained via the Fourier inversion formula
\begin{equation}
  f_{\tilde{\delta}}\brc{x;\delta,\pi_1,\rho}=\frac{1}{2\pi}\int_{-\infty}^{+\infty}e^{-itx}\varphi_{\tilde{\delta}}\brc{t;\delta,\pi_1,\rho}dt.
  \label{eq:pdf}
\end{equation}

Since there is generally no closed-form expression for the inversion integral (\ref{eq:pdf}) based on the characterisitic function in (\ref{eq:cf}), numerical approaches such as the fast Fourier transform (FFT) are employed to evaluate the density on a fine grid. In particular, we adopt the \textit{Fastest Fourier Transform in the West} (FFTW) algorithm~\cite{FFTW.jl-2005} to compute the density function numerically. 



Our interest is to construct a confidence interval for the risk difference $\delta$. A procedure analogous to the Neyman construction is considered. Let $A\brc{\delta,\pi_1,\rho}$ denote the acceptance region for $\tilde{\delta}$ under parameter values $\brc{\delta,\pi_1,\rho}$. Then, a $\brc{1-\alpha}$-level confidence interval for the risk difference $\delta$ consists of all values of $\delta_0$ such that
\begin{equation}
  \mathrm{CI}_{1-\alpha}\brc{\delta}=\left\{\delta_0:~\tilde{\delta}_\text{obs}\in A\brc{\delta_0,\bar{\pi}_1,\bar{\rho}}\right\}, 
\end{equation}
where $\bar{\pi}_1$ and $\bar{\rho}$ denote plug-in estimates of the nuisance parameters.
In particular, the acceptance region $A\brc{\delta_0,\bar{\pi}_1,\bar{\rho}}$ is associated with the interval $\brc{q_{\alpha_1},~q_{1-\alpha_2}}$, where $q_{\alpha_1}$ and $q_{1-\alpha_2}$ denote the $\alpha_1$- and $1-\alpha_2$-quantiles of the cumulative distribution function (CDF) associated with the PDF of $\tilde{\delta}$ conditional on $\brc{\delta_0,\bar{\pi}_1,\bar{\rho}}$, such that $Pr\brc{q_{\alpha_1}<\tilde{\delta}\le q_{1-\alpha_2}\,\mid\,\delta_0,\bar{\pi}_1,\bar{\rho}}=1-\alpha$.

As we shall see in the simulation studies in Section~\ref{sec:results:simulation}, the PDFs are observed to be unimodal across extensive parameter settings. Using the highest density region to construct the shortest interval $\brc{q_{\alpha_1},q_{1-\alpha_2}}$ is analogous to constructing a highest posterior density (HPD) region in Bayesian inference and yet done in a frequentist inversion framework. Specifically, under unimodality, the shortest acceptance region for $\tilde{\delta}$ conditional on $\brc{\delta_0,\bar{\pi}_1,\bar{\rho}}$ is
\begin{equation}
A\brc{\delta_0,\bar{\pi}_1,\bar{\rho}}=\left\{x:~f_{\tilde{\delta}}\brc{x;\delta_0,\bar{\pi}_1,\bar{\rho}}\ge c\brc{\delta_0,\bar{\pi}_1,\bar{\rho}}\right\},
\end{equation}
where
$$
c\brc{\delta_0,\bar{\pi}_1,\bar{\rho}}=f_{\tilde{\delta}}\brc{q_{\alpha_1};\delta_0,\bar{\pi}_1,\bar{\rho}}=f_{\tilde{\delta}}\brc{q_{1-\alpha_2};\delta_0,\bar{\pi}_1,\bar{\rho}}. 
$$

Due to the fact that the sampling distribution of the estimator $\tilde{\delta}$ is centered at the risk difference, increasing (decreasing) $\delta_0$ will shift the PDF for $\tilde{\delta}$ to the right (left). Therefore, an adaptive backtracking search is used to determine the upper and lower bounds of $\delta$ so as to obtain the confidence interval $\mathrm{CI}_{1-\alpha}\brc{\delta}$. For a candidate value $\delta_0$, let $\bar{\pi}_1\brc{\delta_0}$ and $\bar{\rho}\brc{\delta_0}$ denote the plug-in estimates of the nuisance parameters. These may either be fixed at known values or recomputed as profile maximum likelihood estimates (MLEs) under the constraint $\delta=\delta_0$. 
A specific algorithm is outlined as follows. 
\begin{algorithm}
  \caption{Adaptive backtracking search. The algorithm below is for the upper confidence limit. The same algorithm is excuted for the lower confidence limit by reversing the search direction, i.e., replacing $flag\gets1$ with $flag\gets-1$ (and vice versa).}\label{alg:a1}
  \begin{algorithmic}
    \State Initialize: $\delta_0\gets\delta_0^{(0)}$; $step\gets0.1$
    \State Set: $\epsilon\gets10^{-5}$
    \While{$step>\epsilon$}
    \If{$\tilde{\delta}_\text{obs}\in A\brc{\delta_0,\bar{\pi}_1\brc{\delta_0},\bar{\rho}\brc{\delta_0}}$}
    \State $flag\gets1$ \Comment{Search toward larger values of $\delta_0$}
    \Else
    \State $flag\gets-1$ \Comment{Reverse the search direction and halve the step size}
    \State $step\gets step/2$
    \EndIf
    \State $\delta_0\gets\delta_0+flag\times step$
    \EndWhile
    \State $\delta_U\gets\delta_0$ \Comment{Store the upper confidence limit}
  \end{algorithmic}
\end{algorithm}

\subsection{Asymptotic Methods}
\label{sec:methods:asymp}
Three commonly used likelihood-based asymptotic statistics are used to construct the confidence interval for the risk difference $\delta$, which are the Wald-type, likelihood ratio and score tests, respectively. In the recent work by Zhou and Ma~\cite{zhou2025testing}, the performance of these three asymptotic statistics is examined for testing risk difference of two proportions under Donner's constant $\rho$ model. In what follows, we review these asymptotic methods and describe the corresponding confidence interval constructions for $\delta$.

The likelihood-based methods rely on the MLEs for the parameters. Based on the distributions in (\ref{eq:distribution}) and expressions of the cell probabilities in (\ref{eq:prob_ri}), the likelihood may be parameterized by $\bgam=\brc{\delta,\pi_1,\rho}$ after reparameterizing via $\pi_2=\pi_1+\delta$. Let the global MLEs of $\bgam$ be $\hat{\bgam}=\brc{\hat{\delta},\hat{\pi}_1,\hat{\rho}}$, and the profile MLEs obtained under the constraint $\delta=\delta_0$ be $\hat{\bgam}_0=\brc{\hat{\pi}_{1,0},\hat{\rho}_0}$. The global and constrained profile MLEs are obtained numerically via iterative procedures and the details can be found in the related work by Zhou and Ma~\cite{zhou2025testing}.

\subsubsection{Wald-type Test Induced Confidence Interval}
Based on the asymptotic normality of the MLEs that $\hat{\bgam}-\bgam\sim AN\brc{\boldsymbol{0},\bI\brc{\hat{\bgam}}^{-1}}$, the Wald-type test statistic under $H_0:~\delta=\delta_0$ is given by
\begin{equation}
  Q_W=\frac{\hat{\delta}-\delta_0}{\sqrt{\brc{\bI\brc{\hat{\bgam}}^{-1}}_{11}}}\sim AN\brc{0,1},
\end{equation}
where $\brc{\bI\brc{\hat{\bgam}}^{-1}}_{11}$ denotes the $\brc{1,1}$-th element of the inverse of the Fisher information matrix. The expression of the Fisher information $\bI\brc{\bgam}$ matrix as well as the derivation of this quantity based on the elements of the Fisher information can be found in the paper by Zhou and Ma~\cite{zhou2025testing}.

By the asymptotic pivotal argument, a $\brc{1-\alpha}$-level confidence interval for $\delta$ is given by
\begin{equation}
  \mathrm{CI}_{1-\alpha}\brc{\delta}=\left\{\delta_0:~\left|Q_W\brc{\delta_0}\right|\le z_{1-\alpha/2}\right\}.
\end{equation}

\subsubsection{Likelihood Ratio Test Induced Confidence Interval}
Under the null hypothesis $H_0:~\delta=\delta_0$, the likelihood ratio test statistic
\begin{equation}
  Q_{LR}=2\left[\ell_1\brc{\hat{\bgam}}-\ell_1\brc{\delta_0,\hat{\bgam}_0}\right] 
\end{equation}
asymptotically follows a chi-square distribution with one degree of freedom, where $\ell_1\brc{\cdot}$ denotes the log-likelihood function and its explicit form can be found in Zhou and Ma~\cite{zhou2025testing}. Thus, a $\brc{1-\alpha}$-level confidence interval based on the likelihood ratio test is given by
\begin{equation}
  \mathrm{CI}_{1-\alpha}\brc{\delta}=\left\{\delta_0:~Q_{LR}\brc{\delta_0}\le\chi_{1,1-\alpha}^2\right\}.
\end{equation}

Analogous to the distribution-based approach, this confidence interval can be obtained with adaptive backtracking search algorithm outlined in \textbf{Algorithm}~\ref{alg:a1}, where the \textbf{if}-condition is replaced with $Q_{LR}\brc{\delta_0}\le\chi_{1,1-\alpha}^2$.

\subsubsection{Score Test Induced Confidence Interval}
The score test under the null hypothesis $H_0:~\delta=\delta_0$ is given by
\begin{equation}
  Q_S=\left.\bU\brc{\bgam}\bI\brc{\bgam}^{-1}\bU\brc{\bgam}^T\right|_{\bgam=\brc{\delta_0,\hat{\bgam}_0}}=\left.\brc{\bI\brc{\bgam}^{-1}}_{11}\brc{\frac{\partial\ell_1\brc{\bgam}}{\partial\delta}}^2\right|_{\bgam=\brc{\delta_0,\hat{\bgam}_0}}, 
\end{equation}
where $\bU\brc{\bgam}=\nabla_{\bgam}^T\ell_1=\brc{\frac{\partial\ell_1}{\partial\delta},\frac{\partial\ell_1}{\partial\pi_1},\frac{\partial\ell_1}{\partial\rho}}$ is the score function, and $\brc{\bI\brc{\bgam}^{-1}}_{11}$ is the $(1,1)$-th element in the inverse of the Fisher information matrix $\bI\brc{\bgam}$.

The score test statistic asymptotically follows a chi-square distribution with one degree of freedom. Likewise, a $(1-\alpha)$-level confidence interval is constructed by
\begin{equation}
  \mathrm{CI}_{1-\alpha}\brc{\delta}=\left\{\delta_0:~Q_S\brc{\delta_0}\le\chi_{1,1-\alpha}^2\right\}.
\end{equation}

Similarly, this confidence interval can be computed using \textbf{Algorithm}~\ref{alg:a1} with the \textbf{if}-condition replaced with $Q_S\brc{\delta_0}\le\chi_{1,1-\alpha}^2$.

\subsection{MOVER-based Confidence Interval}
\label{sec:methods:mover}
The method of variance estimates recovery (MOVER) was introduced by Zou and Donner~\cite{Zou_2008con} as a general approach to construct confidence intervals for effect measures. Basically, under asymptotic normality, the confidence interval (denoted by $\brc{L,U}$) of a difference between effect measures characterized by $\theta_1-\theta_2$ is given by
\begin{equation}
  \brc{L,U}=\hat{\theta}_1-\hat{\theta}_2\mp z_{1-\alpha/2}\sqrt{\widehat{\var}\brc{\hat{\theta}_1}+\widehat{\var}\brc{\hat{\theta}_2}},
  \label{eq:ci:mover}
\end{equation}
where $\hat{\theta}_i$ and $\widehat{\var}\brc{\hat{\theta}_i}$ are the point estimate and variance estimator for $\theta_i,~i=1,2$.

The MOVER approach aims to improve the performance of the confidence limits in (\ref{eq:ci:mover}) for small to moderate samples by obtaining the variance estimator $\widehat{\var}\brc{\hat{\theta}_i}$ in the neighborhood of the lower and upper limits, respectively. Specifically, let $\brc{l_i,u_i}$ be a $\brc{1-\alpha}$-level confidence interval for $\theta_i$, then applying the inversion principle to the Wald test statistic $T_{w,i}=\brc{\hat{\theta}_i-\theta_i}^2/\widehat{\var}\brc{\hat{\theta}_i}$, the variance estimator can be evaluated with $\theta_i=l_i,u_i$, i.e.,
\begin{equation}
  \widehat{\var}\brc{\hat{\theta}_i}=\frac{\brc{\hat{\theta}_i-l_i}^2}{z_{1-\alpha/2}^2}=\frac{\brc{u_i-\hat{\theta}_i}^2}{z_{1-\alpha/2}^2}.
\end{equation}

Therefore, $\widehat{\var}\brc{\hat{\theta}_1}$ and $\widehat{\var}\brc{\hat{\theta}_2}$ are evaluated with $\theta_1=l_1(u_1)$ and $\theta_2=u_2(l_2)$ when estimating $L(U)$, in the sense that these evaluations are in the vicinity of the limits (i.e., $L\approx l_1-u_2$, $U\approx u_1-l_2$). Thus, we have
\begin{equation}
  \begin{aligned}
    L&=\hat{\theta}_1-\hat{\theta}_2-\sqrt{\brc{\hat{\theta}_1-l_1}^2+\brc{u_2-\hat{\theta}_2}^2}, \\
    U&=\hat{\theta}_1-\hat{\theta}_2+\sqrt{\brc{u_1-\hat{\theta}_1}^2+\brc{\hat{\theta}_2-l_2}^2}. 
  \end{aligned}
\end{equation}

Particularly, referring to the risk difference, we have $\theta_1=\pi_2$ and $\theta_2=\pi_1$. Moreover, in the context of intra-subject correlation, the variance estimates have to be adjusted by introducing an effect factor denoted by $e_i$ such that $\var\brc{\hat{\pi}_i}=\var\brc{\hat{\pi}_i^0}/e_i$, where $\var\brc{\hat{\pi}_i^0}$ is the variance estimate when paired measurements are independent, i.e., $\rho=0$. Conditioning on the total cell counts in each group and using expression in (\ref{eq:pis}) for the point estimates, it is straightforward to show that
\begin{equation}
  \var\brc{\hat{\pi}_i}=\var\brc{\hat{\pi}_i^0}\cdot\frac{2m_{+i}\brc{1+\rho}+n_{+i}}{2m_{+i}+n_{+i}}
  \,\Longrightarrow\,
  e_i=\frac{2m_{+i}+n_{+i}}{2m_{+i}\brc{1+\rho}+n_{+i}}, 
\end{equation}
based on the distributions in (\ref{eq:distribution}) and cell probabilities in (\ref{eq:prob_ri}). Eventually, the confidence interval for $\delta$ is
\begin{equation}
  \begin{aligned}
    \delta_L&=\hat{\pi}_2-\hat{\pi}_1-\sqrt{\frac{\brc{\hat{\pi}_2-l_{\pi_2}}^2}{e_2}+\frac{\brc{u_{\pi_1}-\hat{\pi}_1}^2}{e_1}}, \\
    \delta_U&=\hat{\pi}_2-\hat{\pi}_1+\sqrt{\frac{\brc{u_{\pi_2}-\hat{\pi}_2}^2}{e_2}+\frac{\brc{\hat{\pi}_1-l_{\pi_1}}^2}{e_1}}.
  \end{aligned}
  \label{eq:ci:mover:rd}
\end{equation}

We consider two methods for obtaining the confidence limits $\brc{l_{\pi_i},u_{\pi_i}}$ as follows.
\begin{enumerate}
\item With Wilson score method~\cite{wilson1927probable}, the interval is given by
  \begin{equation}
    \brc{l_{\pi_i},u_{\pi_i}}=\frac{\tilde{\pi}_i+z_{1-\alpha/2}^2/2\tilde{n}_i}{1+z_{1-\alpha/2}^2/\tilde{n}_i}\mp\frac{z_{1-\alpha/2}\sqrt{\tilde{\pi}_i\brc{1-\tilde{\pi}_i}/\tilde{n}_i+z_{1-\alpha/2}^2/4\tilde{n}_i^2}}{1+z_{1-\alpha/2}^2/\tilde{n}_i}, 
  \end{equation}
  where $\tilde{\pi}_i$ takes form in (\ref{eq:pis}), and $\tilde{n}_i=2m_{+i}+n_{+i}$ denoting the total number of responses in the $i$-th group.

\item Alternatively, with Agresti-Coull method~\cite{agresti1998approximate}, the interval is estimated as
  \begin{equation}
    \brc{l_{\pi_i},u_{\pi_i}}=\tilde{p}_i\mp z_{1-\alpha/2}\sqrt{\tilde{p}_i\brc{1-\tilde{p}_i}/\bar{n}_i},
    \label{eq:agresti-coull}
  \end{equation}
  where $\tilde{p}_i=\frac{\tilde{\pi}_i+z_{1-\alpha/2}^2/2\tilde{n}_i}{1+z_{1-\alpha/2}^2/\tilde{n}_i}$ and $\bar{n}_i=\tilde{n}_i+z_{1-\alpha/2}^2$.
\end{enumerate}

Note that the confidence limits in (\ref{eq:ci:mover:rd}) depend on the estimates of $\pi_1$, $\pi_2$, and $\rho$, which are estimated by their MLEs in numerical evaluations.

\section{Results}
\label{sec:results}

\subsection{Simulation Studies}
\label{sec:results:simulation}
We implement simulation studies to assess the performance of the distribution-based approach along with the three commonly used asymptotic methods and MOVER-based methods by investigating the three quantities: (i) the empirical coverage probability (ECP), (ii) the mean interval width (MIW), and (iii) the ratio of mesial non-coverage probability to the distal non-coverage probability (RMNCP), whose formulae for a given CI under $H_0:~\delta=\delta_0$ are expressed respectively as below.
\begin{equation}
  \begin{aligned}
    &\mathrm{ECP}=\frac{1}{N}\sum_{i=1}^NI\brc{\left\{\delta_L^{(i)}\le\delta_0\le\delta_U^{(i)}\right\}}, \\
    &\mathrm{MIW}=\frac{1}{N}\sum_{i=1}^N\brc{\delta_U^{(i)}-\delta_L^{(i)}}, \\
    &\mathrm{RMNCP}=\frac{\sum_{i=1}^NI\brc{\left\{\delta_0<\delta_L^{(i)}\right\}}}{\sum_{i=1}^NI\brc{\left\{\delta_0<\delta_L^{(i)}\right\}\cup\left\{\delta_0>\delta_U^{(i)}\right\}}},
  \end{aligned}
  \label{eq:ecp:miw:rmncp}
\end{equation}
where $N$ is the total number of replications in the Monte Carlo simulation process, and $\delta_L^{(i)}(\delta_U^{(i)})$ the lower (upper) confidence limit for the $i$-th replication.

\subsubsection{Simulation Designs}
We first investigate several representative parameter configurations. Equal bilateral and unilateral sample sized are assumed in both groups, namely $m_{+i}=n_{+i}=20,50,100$ for $i=1,2$, corresponding to small, moderate and large sample size, respectively. For simplicity, this setting is denoted by ``$m=n$'' and is hereafter referred to as the ``equal sample size'' setting. The intra-subject correlation coefficient is set as $\rho=0,0.5,0.9$. The response probability in group 1 is taken as $\pi_1=0.1,0.2,0.3$, and the null risk difference is specified as $\delta_0=0.1,0.2,0.3$. For each parameter configuration $\brc{\rho,\pi_1,m=n,\delta_0}$, $N=10,000$ datasets are generated and the confidence intervals are constructed using the methods described in Section~\ref{sec:methods:dist} -- \ref{sec:methods:mover}. The three quantities (ECP, MIW and RMNCP) are then calculated according to (\ref{eq:ecp:miw:rmncp}).

Additionally, to further assess the robustness of the competing methods across a broad parameter space, we randomly generate $1,000$ parameter configurations while fixing $m=n=20,50,100$. Specifically, the parameters are sampled from $\rho\in\brc{-1,1}$, $\pi_1\in\brc{0,1}$ and $\delta_0\in\brc{0,1}$, subject to the model admissibility constraints such that all implied probabilities are valid, i.e., $\pi_2,p_{ri}\in\brc{0,1}$. For each randomly generated configuration, the ECP, MIW, and RMNCP are evaluated using $N=10,000$ replications. 

In the subsequent tables and figures, the methods are denoted by $Q_W$, $Q_{LR}$, $Q_S$, $\mathrm{MV}_1$, $\mathrm{MV}_2$, $\mathrm{PDF}_1$, and $\mathrm{PDF}_2$. Here, $Q_W$, $Q_{LR}$ and $Q_S$ correspond to the Wald-type, likelihood ratio, and score methods, respectively; $\mathrm{MV}_1$ and $\mathrm{MV}_2$ denote the MOVER procedures based on the Wilson score and Agresti-Coull intervals; and $\mathrm{PDF}_1$ and $\mathrm{PDF}_2$ represent the proposed distribution-based approach using the true values of $\brc{\pi_1,\rho}$ and the profile MLEs, respectively. The CIs constructed in simulations are at 95\%-level, i.e., $\alpha=0.05$.

\subsubsection{Simulation Results for ECP}
Table~\ref{tb:ecp} provides the ECPs based on $N=10,000$ simulations under different parameter configurations designed as above. Values smaller than $0.94$ and greater than $0.96$ are highlighted in the table to indicate liberal and conservative performance, respectively. As can be seen, all methods achieve ECPs reasonably close to the nominal level of $0.95$, with all reported values exceeding $0.90$. Departures from the nominal level occur primarily under the equal sample size setting with $m=n=20$, particularly when the intra-subject correlation is either small ($\rho=0$) or large ($\rho=0.9$), and/or when the baseline probability is small ($\pi_1=0.1$). For example, the score method $Q_S$ tends to be conservative when $\brc{\rho,\pi_1,m=n}=\brc{0,0.1,20}$, whereas $\mathrm{MV}_1$ and $\mathrm{PDF}_1$ exhibit conservative behavior when $\brc{\rho,\pi_1,m=n}=\brc{0.5,0.1,20}$. Under the more challenging setting $\brc{\rho,\pi_1,m=n}=\brc{0.9,0.1,20}$, $\mathrm{PDF}_1$ remains conservative while $\mathrm{PDF}_2$ becomes somewhat liberal. Nevertheless, such departures become substantially less frequent as the sample size increases. 
{\scriptsize
\begin{longtblr}[
   caption={ECP for a 95\%-level CI based on $N=10,000$ simulations across different parameter settings. Note that the values that are smaller than $0.94$ or greater than $0.96$ are highlighted in \textbf{bold}, indicating the liberal and conservative results, respectively.},
   label={tb:ecp}
 ]{
 }
 \toprule
 $\rho$ &$\pi_1$ &$m=n$ &$\delta_0$ &$Q_W$ &$Q_{LR}$ &$Q_S$ &$\mathrm{MV}_1$ &$\mathrm{MV}_2$ &$\mathrm{PDF}_1$ &$\mathrm{PDF}_2$ \\
 \midrule
0.0 &0.1 & 20 &0.1 &\textbf{0.9664} &\textbf{0.9676} &\textbf{0.9705} &\textbf{0.9733} &\textbf{0.9764} &\textbf{0.9806} &\textbf{0.9810} \\
& & &0.2 &0.9563 &\textbf{0.9608} &\textbf{0.9644} &\textbf{0.9633} &\textbf{0.9672} &0.9449 &\textbf{0.9613} \\
& & &0.3 &0.9542 &0.9559 &\textbf{0.9610} &0.9565 &0.9587 &0.9491 &0.9493 \\
& & 50 &0.1 &0.9566 &0.9595 &\textbf{0.9610} &\textbf{0.9609} &\textbf{0.9626} &\textbf{0.9616} &\textbf{0.9615} \\
& & &0.2 &0.9497 &0.9472 &0.9552 &0.9546 &0.9562 &0.9496 &0.9522 \\
& & &0.3 &0.9502 &0.9505 &0.9540 &0.9527 &0.9538 &0.9544 &0.9533 \\
& &100 &0.1 &0.9540 &0.9551 &0.9561 &0.9557 &0.9568 &0.9546 &0.9572 \\
& & &0.2 &0.9503 &0.9495 &0.9532 &0.9538 &0.9549 &0.9510 &0.9524 \\
& & &0.3 &0.9543 &0.9518 &0.9564 &0.9553 &0.9564 &0.9483 &0.9515 \\
&0.2 & 20 &0.1 &0.9550 &0.9578 &0.9590 &0.9590 &\textbf{0.9615} &\textbf{0.9783} &0.9589 \\
& & &0.2 &0.9513 &0.9529 &0.9597 &0.9553 &0.9571 &0.9598 &0.9588 \\
& & &0.3 &0.9448 &\textbf{0.9353} &0.9520 &0.9481 &0.9489 &0.9457 &0.9498 \\
& & 50 &0.1 &0.9472 &0.9492 &0.9509 &0.9505 &0.9512 &0.9488 &0.9509 \\
& & &0.2 &0.9488 &0.9486 &0.9527 &0.9510 &0.9516 &0.9489 &0.9480 \\
& & &0.3 &0.9495 &0.9441 &0.9518 &0.9510 &0.9513 &0.9458 &0.9492 \\
& &100 &0.1 &0.9446 &0.9460 &0.9472 &0.9471 &0.9476 &0.9459 &0.9450 \\
& & &0.2 &0.9500 &0.9513 &0.9528 &0.9528 &0.9533 &0.9539 &0.9505 \\
& & &0.3 &0.9470 &0.9469 &0.9485 &0.9485 &0.9491 &0.9471 &0.9471 \\
&0.3 & 20 &0.1 &0.9419 &0.9484 &0.9530 &0.9477 &0.9489 &0.9546 &0.9514 \\
& & &0.2 &0.9458 &0.9503 &0.9553 &0.9488 &0.9495 &0.9466 &0.9531 \\
& & &0.3 &0.9478 &\textbf{0.9390} &0.9562 &0.9513 &0.9520 &0.9565 &0.9586 \\
& & 50 &0.1 &0.9481 &0.9508 &0.9517 &0.9495 &0.9500 &0.9431 &0.9485 \\
& & &0.2 &0.9487 &0.9512 &0.9528 &0.9511 &0.9514 &0.9475 &0.9476 \\
& & &0.3 &0.9488 &0.9487 &0.9522 &0.9508 &0.9511 &0.9517 &0.9501 \\
& &100 &0.1 &0.9493 &0.9513 &0.9520 &0.9506 &0.9507 &0.9487 &0.9494 \\
& & &0.2 &0.9486 &0.9499 &0.9503 &0.9496 &0.9496 &0.9459 &0.9478 \\
& & &0.3 &0.9486 &0.9490 &0.9494 &0.9491 &0.9491 &0.9464 &0.9479 \\
0.5 &0.1 & 20 &0.1 &0.9583 &\textbf{0.9607} &\textbf{0.9648} &\textbf{0.9696} &\textbf{0.9739} &\textbf{0.9719} &\textbf{0.9674} \\
& & &0.2 &0.9463 &0.9516 &0.9547 &0.9583 &\textbf{0.9632} &\textbf{0.9701} &0.9545 \\
& & &0.3 &0.9501 &0.9555 &0.9574 &0.9598 &\textbf{0.9626} &\textbf{0.9601} &0.9545 \\
& & 50 &0.1 &0.9491 &0.9511 &0.9531 &0.9581 &\textbf{0.9601} &\textbf{0.9608} &0.9520 \\
& & &0.2 &0.9483 &0.9491 &0.9501 &0.9538 &0.9553 &0.9490 &0.9469 \\
& & &0.3 &0.9447 &0.9468 &0.9487 &0.9509 &0.9525 &0.9495 &0.9465 \\
& &100 &0.1 &0.9501 &0.9506 &0.9511 &0.9547 &0.9556 &0.9525 &0.9499 \\
& & &0.2 &0.9499 &0.9502 &0.9505 &0.9533 &0.9541 &0.9506 &0.9496 \\
& & &0.3 &0.9488 &0.9497 &0.9498 &0.9525 &0.9528 &0.9498 &0.9496 \\
&0.2 & 20 &0.1 &0.9449 &0.9490 &0.9515 &0.9522 &0.9542 &\textbf{0.9713} &0.9471 \\
& & &0.2 &0.9495 &0.9529 &0.9556 &0.9569 &0.9598 &\textbf{0.9666} &0.9536 \\
& & &0.3 &0.9417 &0.9454 &0.9481 &0.9482 &0.9499 &\textbf{0.9375} &0.9440 \\
& & 50 &0.1 &0.9447 &0.9459 &0.9469 &0.9514 &0.9531 &0.9438 &0.9477 \\
& & &0.2 &0.9479 &0.9497 &0.9508 &0.9537 &0.9546 &0.9507 &0.9505 \\
& & &0.3 &0.9465 &0.9488 &0.9504 &0.9538 &0.9544 &0.9490 &0.9463 \\
& &100 &0.1 &0.9487 &0.9493 &0.9495 &0.9516 &0.9522 &0.9464 &0.9466 \\
& & &0.2 &0.9503 &0.9503 &0.9505 &0.9533 &0.9534 &0.9482 &0.9508 \\
& & &0.3 &0.9454 &0.9461 &0.9466 &0.9496 &0.9500 &0.9473 &0.9464 \\
&0.3 & 20 &0.1 &0.9436 &0.9486 &0.9510 &0.9510 &0.9516 &\textbf{0.9638} &0.9442 \\
& & &0.2 &0.9459 &0.9510 &0.9527 &0.9515 &0.9527 &0.9474 &0.9489 \\
& & &0.3 &0.9413 &0.9471 &0.9502 &0.9494 &0.9512 &\textbf{0.9720} &0.9432 \\
& & 50 &0.1 &0.9478 &0.9504 &0.9509 &0.9514 &0.9516 &0.9480 &0.9463 \\
& & &0.2 &0.9487 &0.9510 &0.9510 &0.9528 &0.9531 &0.9481 &0.9490 \\
& & &0.3 &0.9459 &0.9457 &0.9492 &0.9509 &0.9511 &0.9479 &0.9463 \\
& &100 &0.1 &0.9483 &0.9496 &0.9503 &0.9504 &0.9504 &0.9470 &0.9472 \\
& & &0.2 &0.9499 &0.9509 &0.9503 &0.9530 &0.9532 &0.9512 &0.9500 \\
& & &0.3 &0.9488 &0.9496 &0.9505 &0.9533 &0.9535 &0.9527 &0.9505 \\
0.9 &0.1 & 20 &0.1 &0.9536 &0.9593 &\textbf{0.9636} &\textbf{0.9655} &\textbf{0.9720} &\textbf{0.9606} &\textbf{0.9185} \\
& & &0.2 &0.9479 &0.9543 &0.9591 &0.9583 &\textbf{0.9623} &\textbf{0.9607} &\textbf{0.9324} \\
& & &0.3 &0.9464 &0.9513 &0.9532 &0.9538 &0.9565 &\textbf{0.9613} &\textbf{0.9111} \\
& & 50 &0.1 &0.9475 &0.9476 &0.9498 &0.9524 &0.9553 &\textbf{0.9739} &0.9403 \\
& & &0.2 &0.9491 &0.9502 &0.9518 &0.9524 &0.9552 &0.9495 &0.9412 \\
& & &0.3 &0.9462 &0.9482 &0.9489 &0.9498 &0.9519 &0.9437 &\textbf{0.9395} \\
& &100 &0.1 &0.9486 &0.9492 &0.9504 &0.9544 &0.9555 &0.9532 &0.9484 \\
& & &0.2 &0.9513 &0.9521 &0.9525 &0.9539 &0.9547 &0.9473 &0.9498 \\
& & &0.3 &0.9492 &0.9499 &0.9504 &0.9527 &0.9532 &0.9502 &0.9474 \\
&0.2 & 20 &0.1 &0.9486 &0.9534 &0.9563 &0.9586 &\textbf{0.9615} &\textbf{0.9720} &\textbf{0.9125} \\
& & &0.2 &0.9457 &0.9511 &0.9526 &0.9553 &0.9571 &\textbf{0.9711} &\textbf{0.9252} \\
& & &0.3 &\textbf{0.9396} &0.9455 &0.9475 &0.9477 &0.9499 &0.9447 &\textbf{0.9085} \\
& & 50 &0.1 &0.9479 &0.9491 &0.9498 &0.9538 &0.9551 &0.9522 &0.9444 \\
& & &0.2 &0.9533 &0.9553 &0.9560 &0.9564 &0.9568 &0.9510 &0.9473 \\
& & &0.3 &0.9460 &0.9489 &0.9491 &0.9540 &0.9546 &0.9451 &0.9436 \\
& &100 &0.1 &0.9474 &0.9479 &0.9484 &0.9548 &0.9552 &0.9489 &0.9474 \\
& & &0.2 &0.9476 &0.9482 &0.9482 &0.9552 &0.9557 &0.9494 &0.9499 \\
& & &0.3 &0.9476 &0.9491 &0.9490 &0.9511 &0.9511 &0.9470 &0.9454 \\
&0.3 & 20 &0.1 &0.9476 &0.9520 &0.9530 &0.9577 &0.9587 &\textbf{0.9768} &\textbf{0.9199} \\
& & &0.2 &\textbf{0.9390} &0.9419 &0.9455 &0.9509 &0.9519 &0.9508 &\textbf{0.9261} \\
& & &0.3 &0.9454 &0.9416 &0.9496 &0.9512 &0.9518 &\textbf{0.9741} &\textbf{0.9164} \\
& & 50 &0.1 &0.9464 &0.9479 &0.9483 &0.9537 &0.9541 &0.9486 &0.9478 \\
& & &0.2 &0.9445 &0.9463 &0.9471 &0.9509 &0.9512 &0.9479 &0.9439 \\
& & &0.3 &0.9458 &0.9459 &0.9472 &0.9508 &0.9510 &0.9447 &0.9426 \\
& &100 &0.1 &0.9486 &0.9494 &0.9497 &0.9526 &0.9526 &0.9470 &0.9475 \\
& & &0.2 &0.9462 &0.9479 &0.9482 &0.9522 &0.9523 &0.9469 &0.9476 \\
& & &0.3 &0.9512 &0.9524 &0.9533 &0.9541 &0.9542 &0.9487 &0.9485 \\
 \bottomrule
\end{longtblr}
}

\begin{figure}[htpb]
  \centering
  \includegraphics[scale=0.36]{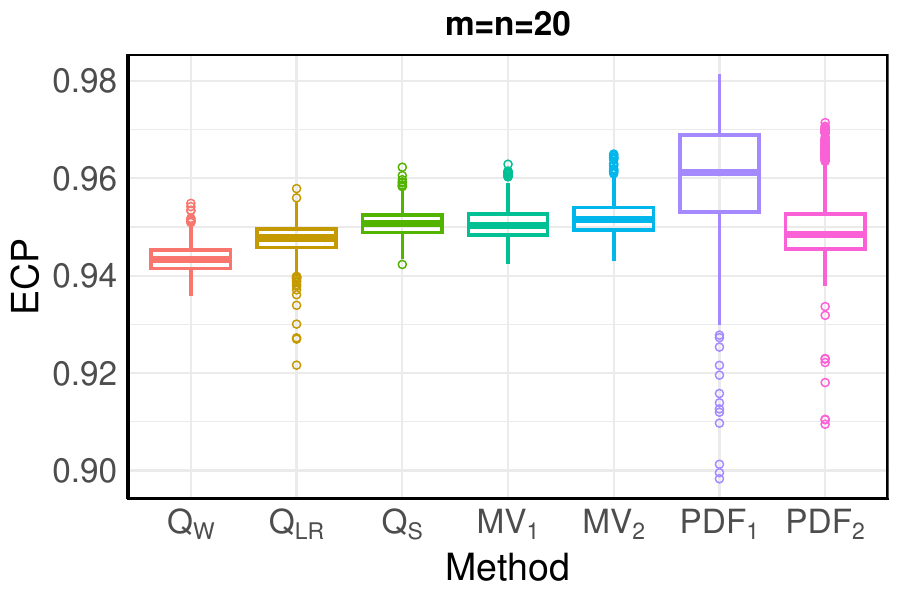}
  \includegraphics[scale=0.36]{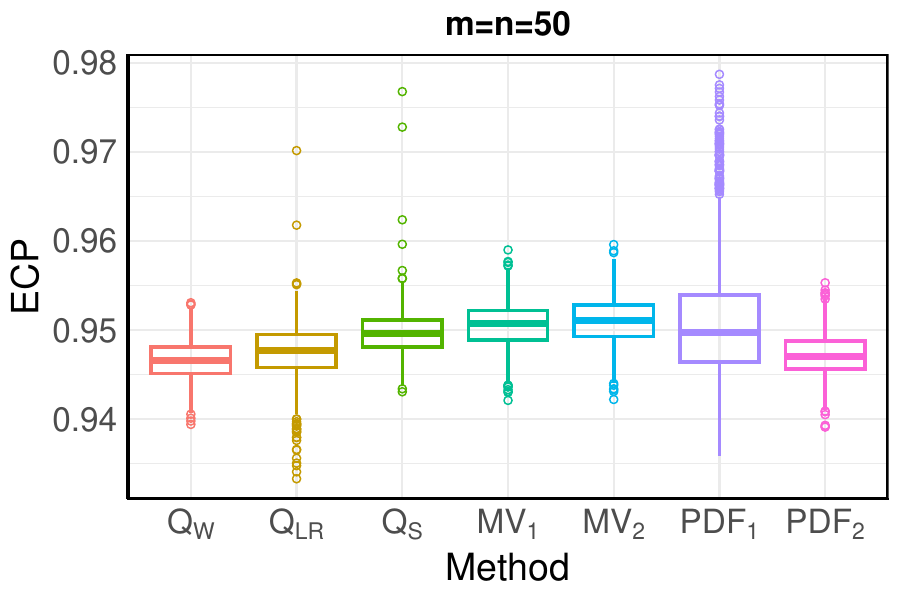}
  \includegraphics[scale=0.36]{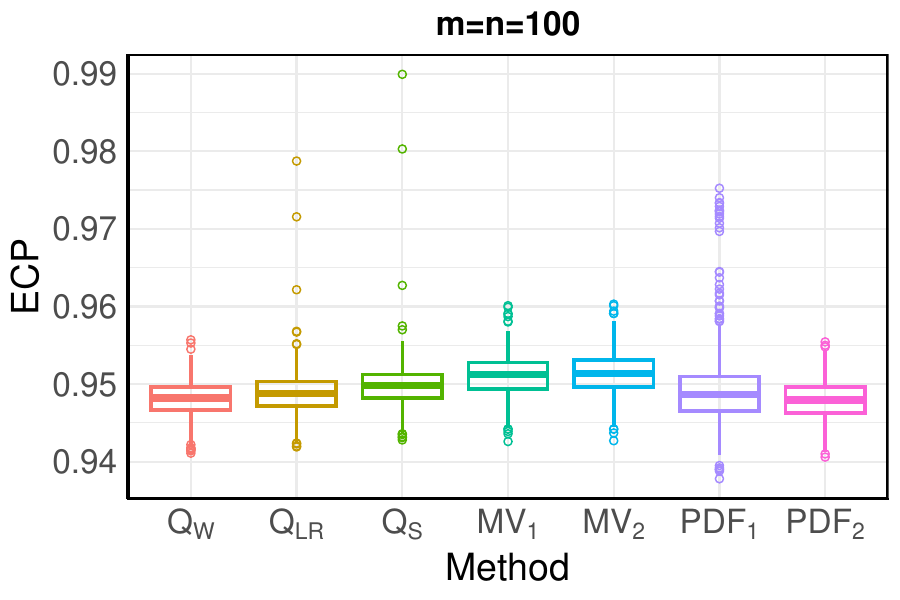}
  \caption{Box plots of the empirical coverage probability (ECP) for the three asymptotic methods ($Q_W$, $Q_{LR}$, and $Q_S$), MOVER procedures ($\mathrm{MV_1}$ and $\mathrm{MV}_2$), and the distribution-base methods ($\mathrm{PDF}_1$ and $\mathrm{PDF}_2$). The panels correspond to $m=n=20,50$ and $100$, respectively. Results are based on $1,000$ randomly generated admissible parameter configurations.}
  \label{fig:boxplot:ecp}
\end{figure}

Figure~\ref{fig:boxplot:ecp} displays box plots of the ECPs obtained from $1,000$ randomly generated admissible parameter configurations for $m=n=20,50$ and $100$. When $m=n=20$, $Q_S$ exhibits a median closest to the nominal level of $0.95$ together with the narrowest interquartile range, indicating comparatively stable coverage performance across the parameter space. This is consistent with the finding in the previous study by Zhou and Ma~\cite{zhou2025testing}. In contrast, $\mathrm{PDF}_1$ has the largest median ECP (approximately $0.96$) and the widest interquartile range, suggesting greater variability and a tendency toward conservative coverage. As the sample size increases, the ECP distributions associated with $\mathrm{PDF}_1$ and $\mathrm{PDF}_2$ become increasingly concentrated around the nominal level. When $m=n=100$, the box plots for all methods are highly similar, indicating that difference in coverage performance largely vanish in large samples. 

\subsubsection{Simulation Results for MIW}
Table~\ref{tb:miw} presents the MIWs based on $N=10,000$ simulations across designed parameter configurations.
Across all methods, MIWs decrease substantially as the sample size increases and tend to increase with both the baseline probability $\pi_1$ and the intra-subject correlation coefficient $\rho$. In contrast, the effect of the null risk difference $\delta_0$ is relatively modest. Specifically, for fixed $\brc{\rho,\pi_1,m=n}$, the MIWs increase slightly with $\delta_0$ when $\pi_1=0.1$ or $0.2$, whereas a mild non-monotonic pattern is observed when $\pi_1=0.3$.

Comparisons among methods reveal that the distribution-based procedures generally produce shorter confidence intervals. In particular, when $\rho=0$ or $0.5$, $\mathrm{PDF}_1$ typically yields the smallest MIW for a given parameter configuration, while $\mathrm{PDF}_2$ often provides the second smallest MIW. There findings suggest that the proposed distribution-based approach can achieve competitive coverage performance while maintaining relatively narrow confidence intervals. 
{\scriptsize
\begin{longtblr}[
  caption={MIW for a 95\%-level CI based on $N=10,000$ simulations across different parameter settings.},
   label={tb:miw}
 ]{
 }
 \toprule
 $\rho$ &$\pi_1$ &$m=n$ &$\delta_0$ &$Q_W$ &$Q_{LR}$ &$Q_S$ &$\mathrm{MV}_1$ &$\mathrm{MV}_2$ &$\mathrm{PDF}_1$ &$\mathrm{PDF}_2$ \\
 \midrule
0.0 &0.1 & 20 &0.1 &0.2626 &0.2687 &0.2744 &0.2705 &0.2768 &0.2409 &0.2723 \\
& & &0.2 &0.2844 &0.2879 &0.2919 &0.2875 &0.2922 &0.2611 &0.2885 \\
& & &0.3 &0.2981 &0.2993 &0.3019 &0.2981 &0.3019 &0.2749 &0.2999 \\
& & 50 &0.1 &0.1615 &0.1635 &0.1660 &0.1638 &0.1657 &0.1556 &0.1631 \\
& & &0.2 &0.1768 &0.1774 &0.1799 &0.1777 &0.1791 &0.1702 &0.1772 \\
& & &0.3 &0.1855 &0.1857 &0.1872 &0.1856 &0.1867 &0.1782 &0.1851 \\
& &100 &0.1 &0.1130 &0.1141 &0.1150 &0.1140 &0.1147 &0.1111 &0.1133 \\
& & &0.2 &0.1241 &0.1244 &0.1255 &0.1245 &0.1250 &0.1218 &0.1238 \\
& & &0.3 &0.1302 &0.1301 &0.1311 &0.1303 &0.1307 &0.1281 &0.1296 \\
&0.2 & 20 &0.1 &0.3068 &0.3101 &0.3139 &0.3052 &0.3081 &0.2936 &0.3090 \\
& & &0.2 &0.3186 &0.3189 &0.3239 &0.3144 &0.3166 &0.3028 &0.3185 \\
& & &0.3 &0.3232 &0.3192 &0.3261 &0.3175 &0.3195 &0.3069 &0.3234 \\
& & 50 &0.1 &0.1929 &0.1942 &0.1957 &0.1929 &0.1938 &0.1896 &0.1928 \\
& & &0.2 &0.2011 &0.2011 &0.2031 &0.2002 &0.2009 &0.1968 &0.2000 \\
& & &0.3 &0.2037 &0.2027 &0.2053 &0.2024 &0.2030 &0.1997 &0.2022 \\
& &100 &0.1 &0.1370 &0.1375 &0.1380 &0.1370 &0.1373 &0.1355 &0.1366 \\
& & &0.2 &0.1425 &0.1428 &0.1433 &0.1423 &0.1425 &0.1407 &0.1418 \\
& & &0.3 &0.1443 &0.1443 &0.1448 &0.1438 &0.1440 &0.1426 &0.1433 \\
&0.3 & 20 &0.1 &0.3334 &0.3368 &0.3399 &0.3272 &0.3284 &0.3232 &0.3320 \\
& & &0.2 &0.3375 &0.3374 &0.3419 &0.3307 &0.3317 &0.3259 &0.3349 \\
& & &0.3 &0.3345 &0.3295 &0.3373 &0.3310 &0.3322 &0.3231 &0.3328 \\
& & 50 &0.1 &0.2129 &0.2137 &0.2145 &0.2113 &0.2116 &0.2097 &0.2113 \\
& & &0.2 &0.2155 &0.2160 &0.2165 &0.2135 &0.2138 &0.2117 &0.2135 \\
& & &0.3 &0.2130 &0.2126 &0.2136 &0.2135 &0.2138 &0.2096 &0.2109 \\
& &100 &0.1 &0.1512 &0.1514 &0.1516 &0.1506 &0.1507 &0.1496 &0.1503 \\
& & &0.2 &0.1529 &0.1530 &0.1531 &0.1522 &0.1523 &0.1513 &0.1518 \\
& & &0.3 &0.1512 &0.1512 &0.1513 &0.1510 &0.1511 &0.1496 &0.1501 \\
0.5 &0.1 & 20 &0.1 &0.2816 &0.2890 &0.2953 &0.2995 &0.3070 &0.2845 &0.3005 \\
& & &0.2 &0.3095 &0.3128 &0.3157 &0.3218 &0.3274 &0.3029 &0.3197 \\
& & &0.3 &0.3253 &0.3261 &0.3267 &0.3341 &0.3386 &0.3149 &0.3302 \\
& & 50 &0.1 &0.1776 &0.1797 &0.1815 &0.1862 &0.1884 &0.1797 &0.1855 \\
& & &0.2 &0.1955 &0.1964 &0.1972 &0.2022 &0.2039 &0.1963 &0.2005 \\
& & &0.3 &0.2059 &0.2061 &0.2063 &0.2115 &0.2128 &0.2053 &0.2091 \\
& &100 &0.1 &0.1260 &0.1267 &0.1274 &0.1313 &0.1321 &0.1284 &0.1305 \\
& & &0.2 &0.1388 &0.1391 &0.1394 &0.1434 &0.1440 &0.1408 &0.1422 \\
& & &0.3 &0.1460 &0.1461 &0.1462 &0.1501 &0.1506 &0.1472 &0.1487 \\
&0.2 & 20 &0.1 &0.3409 &0.3429 &0.3442 &0.3478 &0.3515 &0.3376 &0.3483 \\
& & &0.2 &0.3558 &0.3558 &0.3555 &0.3594 &0.3621 &0.3470 &0.3582 \\
& & &0.3 &0.3614 &0.3601 &0.3591 &0.3634 &0.3658 &0.3510 &0.3612 \\
& & 50 &0.1 &0.2176 &0.2182 &0.2185 &0.2231 &0.2241 &0.2185 &0.2220 \\
& & &0.2 &0.2270 &0.2270 &0.2269 &0.2314 &0.2322 &0.2272 &0.2299 \\
& & &0.3 &0.2301 &0.2298 &0.2295 &0.2339 &0.2345 &0.2298 &0.2320 \\
& &100 &0.1 &0.1544 &0.1546 &0.1547 &0.1586 &0.1590 &0.1562 &0.1577 \\
& & &0.2 &0.1611 &0.1611 &0.1611 &0.1648 &0.1651 &0.1627 &0.1637 \\
& & &0.3 &0.1634 &0.1633 &0.1632 &0.1667 &0.1669 &0.1646 &0.1654 \\
&0.3 & 20 &0.1 &0.3783 &0.3770 &0.3755 &0.3787 &0.3801 &0.3692 &0.3781 \\
& & &0.2 &0.3827 &0.3805 &0.3786 &0.3822 &0.3834 &0.3724 &0.3803 \\
& & &0.3 &0.3787 &0.3760 &0.3743 &0.3781 &0.3796 &0.3685 &0.3755 \\
& & 50 &0.1 &0.2414 &0.2410 &0.2406 &0.2449 &0.2452 &0.2410 &0.2435 \\
& & &0.2 &0.2445 &0.2439 &0.2434 &0.2476 &0.2479 &0.2436 &0.2458 \\
& & &0.3 &0.2418 &0.2406 &0.2406 &0.2485 &0.2489 &0.2410 &0.2427 \\
& &100 &0.1 &0.1712 &0.1710 &0.1709 &0.1745 &0.1746 &0.1724 &0.1734 \\
& & &0.2 &0.1733 &0.1731 &0.1729 &0.1763 &0.1764 &0.1744 &0.1751 \\
& & &0.3 &0.1715 &0.1713 &0.1711 &0.1747 &0.1749 &0.1724 &0.1731 \\
0.9 &0.1 & 20 &0.1 &0.2972 &0.3066 &0.3135 &0.3225 &0.3306 &0.3120 &0.3020 \\
& & &0.2 &0.3289 &0.3323 &0.3352 &0.3486 &0.3546 &0.3302 &0.3159 \\
& & &0.3 &0.3457 &0.3461 &0.3466 &0.3619 &0.3669 &0.3425 &0.3286 \\
& & 50 &0.1 &0.1908 &0.1930 &0.1952 &0.2045 &0.2069 &0.1970 &0.2009 \\
& & &0.2 &0.2096 &0.2105 &0.2114 &0.2222 &0.2239 &0.2148 &0.2165 \\
& & &0.3 &0.2205 &0.2206 &0.2207 &0.2323 &0.2337 &0.2253 &0.2266 \\
& &100 &0.1 &0.1355 &0.1363 &0.1371 &0.1445 &0.1454 &0.1406 &0.1430 \\
& & &0.2 &0.1489 &0.1492 &0.1495 &0.1577 &0.1584 &0.1538 &0.1558 \\
& & &0.3 &0.1564 &0.1564 &0.1564 &0.1651 &0.1657 &0.1616 &0.1628 \\
&0.2 & 20 &0.1 &0.3633 &0.3650 &0.3658 &0.3796 &0.3835 &0.3682 &0.3589 \\
& & &0.2 &0.3787 &0.3780 &0.3770 &0.3924 &0.3954 &0.3774 &0.3639 \\
& & &0.3 &0.3841 &0.3817 &0.3800 &0.3970 &0.3997 &0.3812 &0.3670 \\
& & 50 &0.1 &0.2331 &0.2335 &0.2338 &0.2455 &0.2466 &0.2384 &0.2418 \\
& & &0.2 &0.2428 &0.2425 &0.2423 &0.2547 &0.2555 &0.2481 &0.2501 \\
& & &0.3 &0.2460 &0.2454 &0.2449 &0.2577 &0.2584 &0.2510 &0.2528 \\
& &100 &0.1 &0.1654 &0.1655 &0.1657 &0.1747 &0.1751 &0.1710 &0.1727 \\
& & &0.2 &0.1722 &0.1721 &0.1721 &0.1813 &0.1816 &0.1780 &0.1791 \\
& & &0.3 &0.1745 &0.1743 &0.1741 &0.1836 &0.1838 &0.1800 &0.1812 \\
&0.3 & 20 &0.1 &0.4022 &0.3996 &0.3972 &0.4137 &0.4153 &0.4007 &0.3963 \\
& & &0.2 &0.4070 &0.4029 &0.4004 &0.4198 &0.4211 &0.4041 &0.3913 \\
& & &0.3 &0.4027 &0.3956 &0.3956 &0.4253 &0.4269 &0.4007 &0.3852 \\
& & 50 &0.1 &0.2579 &0.2572 &0.2566 &0.2696 &0.2700 &0.2631 &0.2659 \\
& & &0.2 &0.2607 &0.2598 &0.2589 &0.2721 &0.2725 &0.2659 &0.2676 \\
& & &0.3 &0.2581 &0.2566 &0.2561 &0.2720 &0.2724 &0.2631 &0.2645 \\
& &100 &0.1 &0.1829 &0.1826 &0.1824 &0.1922 &0.1923 &0.1884 &0.1899 \\
& & &0.2 &0.1850 &0.1847 &0.1844 &0.1943 &0.1944 &0.1906 &0.1918 \\
& & &0.3 &0.1830 &0.1826 &0.1823 &0.1923 &0.1924 &0.1884 &0.1895 \\
 \bottomrule
\end{longtblr}
}

\begin{figure}[htpb]
  \centering
  \includegraphics[scale=0.36]{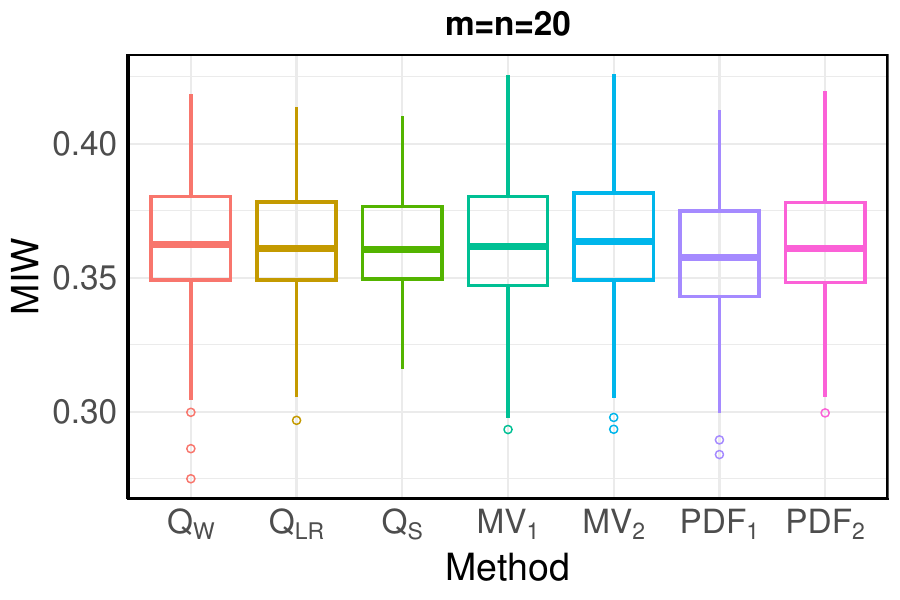}
  \includegraphics[scale=0.36]{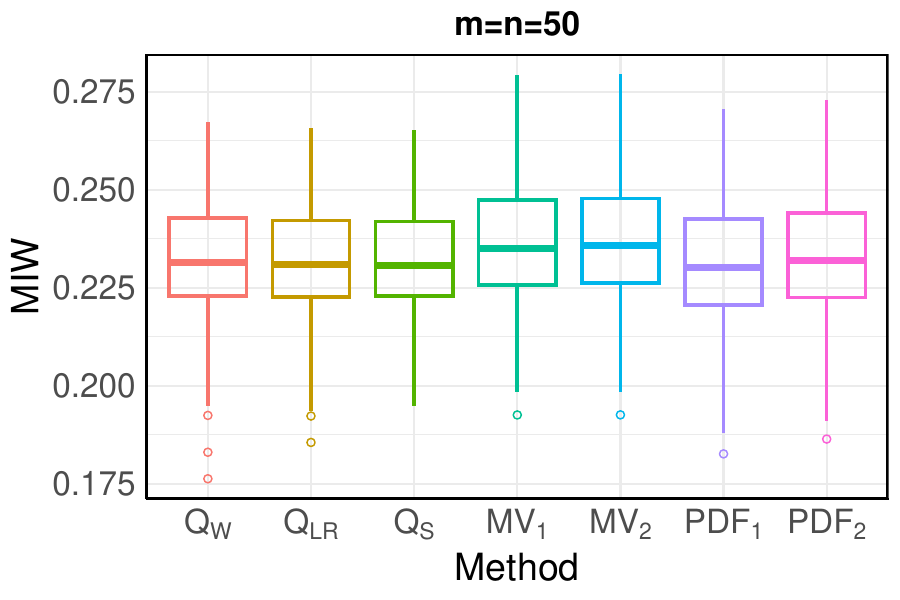}
  \includegraphics[scale=0.36]{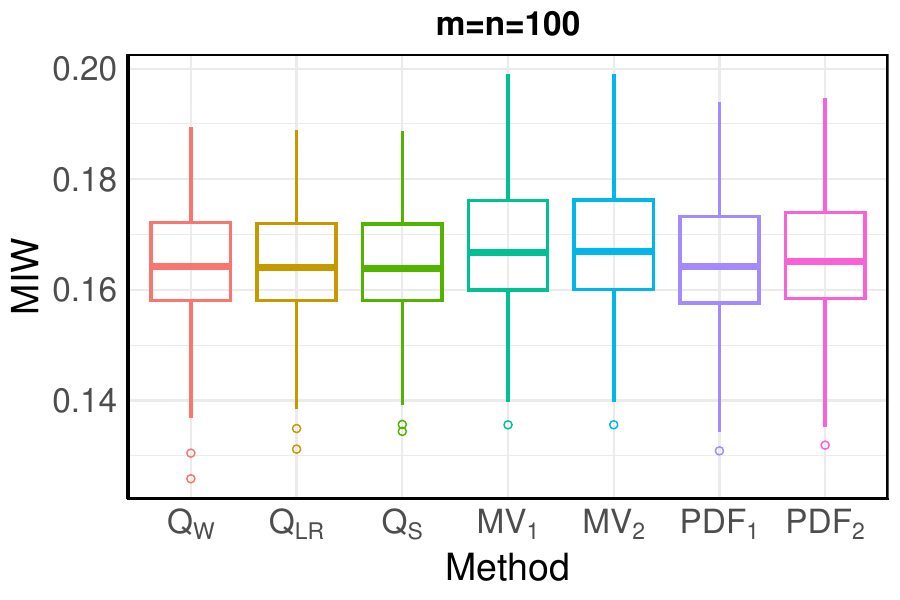}
  \caption{Box plots of the mean interval width (MIW) for methods under consideration. The panels correspond to $m=n=20,50$ and $100$, respectively.}
  \label{fig:boxplot:miw}
\end{figure}

Figure~\ref{fig:boxplot:miw} shows the box plots of the MIWs obtained from $1,000$ randomly generated admissible parameter configurations for $m=n=20,50$ and $100$. Overall, the variability of the MIWs is comparable across methods, and the median MIW decreases as the sample size increases. Among all methods, $\mathrm{PDF}_1$ consistently exhibits the smallest median MIW, indicating a tendency to produce shorter confidence intervals. The advantage becomes less pronounced as the sample size increases. When $m=n=100$, the MIW distributions of $\mathrm{PDF}_1$, $\mathrm{PDF}_2$ and the three asymptotic methods are largely comparable. The MOVER precedures, however, consistently produce slightly larger median MIWs, with the discrepancy becoming marginally more evident as the sample size increases. 

\subsubsection{Simulation Results for RMNCP}
Table~\ref{tb:rmn} presents the RMNCPs based on $N=10,000$ simulations under the designed parameter configurations. Overall, the RMNCPs tend to move toward $0.5$ as the sample size increases. Under challenging settings with small sample size and/or low baseline probability (e.g., $m=n=20$, $\pi_1=0.1$), the asymptotic and MOVER procedures typically yield RMNCP values below $0.5$, indicating a greater proportion of noncoverage occurring in the upper tail. In contrast, the distribution-based procedures occasionally produce extreme RMNCP values, with $\mathrm{PDF}_1$ attaining values of $0$ or $1$ under some configurations.

Recall that the distribution-based procedures construct HPD-like confidence intervals from the recovered finite-sample distribution. Consequently, RMNCP values departing from $0.5$ may reflect asymmetry in the distribution of the risk difference estimator under certain small sample parameter configurations. In contrast, the asymptotic and MOVER procedures are based on asymptotic normality and therefore tend to produce more balanced tail allocations corresponding to a symmetric distributional framework. 
{\scriptsize
\begin{longtblr}[
  caption={RMNCP for a 95\%-level CI based on $10,000$ simulations across different parameter settings.},
   label={tb:rmn}
 ]{
 }
 \toprule
 $\rho$ &$\pi_1$ &$m=n$ &$\delta_0$ &$Q_W$ &$Q_{LR}$ &$Q_S$ &$\mathrm{MV}_1$ &$\mathrm{MV}_2$ &$\mathrm{PDF}_1$ &$\mathrm{PDF}_2$ \\
 \midrule
0.0 &0.1 & 20 &0.1 &0.3363 &0.3684 &0.3797 &0.3071 &0.3178 &1.0000 &0.3413 \\
& & &0.2 &0.2654 &0.2737 &0.2845 &0.1853 &0.1799 &0.2190 &0.1392 \\
& & &0.3 &0.2795 &0.2557 &0.2610 &0.1793 &0.1719 &0.2456 &0.0887 \\
& & 50 &0.1 &0.3664 &0.3918 &0.4064 &0.3760 &0.3717 &0.5864 &0.3631 \\
& & &0.2 &0.3539 &0.3340 &0.3799 &0.3172 &0.3059 &0.4861 &0.3302 \\
& & &0.3 &0.4317 &0.3992 &0.4189 &0.3552 &0.3571 &0.5066 &0.3936 \\
& &100 &0.1 &0.4174 &0.4505 &0.4578 &0.4424 &0.4375 &0.3987 &0.4380 \\
& & &0.2 &0.3944 &0.3758 &0.4013 &0.3745 &0.3769 &0.4306 &0.3929 \\
& & &0.3 &0.4880 &0.4517 &0.4814 &0.4385 &0.4450 &0.5590 &0.4835 \\
&0.2 & 20 &0.1 &0.4200 &0.4177 &0.4055 &0.3780 &0.3688 &0.8291 &0.2976 \\
& & &0.2 &0.4086 &0.4157 &0.4349 &0.3311 &0.3287 &0.4650 &0.2620 \\
& & &0.3 &0.4475 &0.3754 &0.3978 &0.3083 &0.3053 &0.3923 &0.1075 \\
& & 50 &0.1 &0.4867 &0.4841 &0.4877 &0.4707 &0.4672 &0.4902 &0.5105 \\
& & &0.2 &0.5098 &0.4813 &0.4904 &0.4490 &0.4463 &0.5734 &0.4843 \\
& & &0.3 &0.5228 &0.4475 &0.4702 &0.4184 &0.4168 &0.5351 &0.4586 \\
& &100 &0.1 &0.4946 &0.4889 &0.4905 &0.4877 &0.4847 &0.4898 &0.4863 \\
& & &0.2 &0.5360 &0.5041 &0.5064 &0.4852 &0.4818 &0.5054 &0.5213 \\
& & &0.3 &0.5623 &0.5152 &0.5157 &0.4932 &0.4892 &0.5236 &0.5370 \\
&0.3 & 20 &0.1 &0.4888 &0.5043 &0.4883 &0.4359 &0.4305 &0.4365 &0.3704 \\
& & &0.2 &0.5074 &0.4674 &0.4771 &0.4023 &0.3980 &0.3577 &0.3096 \\
& & &0.3 &0.5182 &0.3952 &0.4197 &0.3409 &0.3333 &0.1694 &0.1425 \\
& & 50 &0.1 &0.5279 &0.5136 &0.5106 &0.5050 &0.5040 &0.5079 &0.5272 \\
& & &0.2 &0.5322 &0.4927 &0.4805 &0.4540 &0.4506 &0.3962 &0.4589 \\
& & &0.3 &0.5382 &0.4764 &0.4736 &0.4309 &0.4335 &0.5280 &0.4889 \\
& &100 &0.1 &0.5464 &0.5309 &0.5261 &0.5142 &0.5152 &0.5322 &0.5158 \\
& & &0.2 &0.5467 &0.5161 &0.5000 &0.4901 &0.4901 &0.4991 &0.5211 \\
& & &0.3 &0.5700 &0.5363 &0.5188 &0.4951 &0.4951 &0.5205 &0.5374 \\
0.5 &0.1 & 20 &0.1 &0.3398 &0.3760 &0.4029 &0.3102 &0.3077 &1.0000 &0.3715 \\
& & &0.2 &0.3669 &0.3822 &0.3885 &0.3046 &0.3043 &0.5919 &0.3524 \\
& & &0.3 &0.4549 &0.4247 &0.4131 &0.3159 &0.3128 &0.3990 &0.3510 \\
& & 50 &0.1 &0.4754 &0.5072 &0.5224 &0.4702 &0.4687 &0.6796 &0.5104 \\
& & &0.2 &0.4507 &0.4676 &0.4729 &0.4026 &0.3937 &0.5373 &0.4821 \\
& & &0.3 &0.4991 &0.4831 &0.4717 &0.4297 &0.4274 &0.5762 &0.5121 \\
& &100 &0.1 &0.4529 &0.4818 &0.4888 &0.4459 &0.4459 &0.5127 &0.4890 \\
& & &0.2 &0.5150 &0.5241 &0.5273 &0.5054 &0.5054 &0.5405 &0.5496 \\
& & &0.3 &0.4766 &0.4732 &0.4701 &0.4484 &0.4470 &0.4761 &0.4980 \\
&0.2 & 20 &0.1 &0.5136 &0.5020 &0.4928 &0.4623 &0.4607 &1.0000 &0.4470 \\
& & &0.2 &0.5208 &0.4841 &0.4550 &0.4200 &0.4179 &0.4848 &0.4181 \\
& & &0.3 &0.5540 &0.4927 &0.4509 &0.4035 &0.3972 &0.4375 &0.3781 \\
& & 50 &0.1 &0.5154 &0.5102 &0.5047 &0.4835 &0.4861 &0.5089 &0.5143 \\
& & &0.2 &0.5317 &0.5089 &0.4939 &0.4795 &0.4758 &0.4787 &0.4990 \\
& & &0.3 &0.5776 &0.5195 &0.4819 &0.4524 &0.4496 &0.4784 &0.5140 \\
& &100 &0.1 &0.4620 &0.4596 &0.4594 &0.4669 &0.4644 &0.4701 &0.4607 \\
& & &0.2 &0.4889 &0.4728 &0.4646 &0.4561 &0.4571 &0.4730 &0.4675 \\
& & &0.3 &0.5366 &0.5065 &0.4906 &0.4464 &0.4440 &0.4915 &0.4851 \\
&0.3 & 20 &0.1 &0.5035 &0.4650 &0.4469 &0.4571 &0.4566 &0.7239 &0.4355 \\
& & &0.2 &0.5767 &0.5092 &0.4852 &0.4701 &0.4630 &0.4819 &0.5039 \\
& & &0.3 &0.6474 &0.5589 &0.5162 &0.4881 &0.4734 &0.0000 &0.4760 \\
& & 50 &0.1 &0.4981 &0.4718 &0.4623 &0.4650 &0.4628 &0.4885 &0.4804 \\
& & &0.2 &0.5458 &0.5020 &0.4816 &0.4725 &0.4712 &0.4721 &0.4667 \\
& & &0.3 &0.5648 &0.4733 &0.4615 &0.4420 &0.4397 &0.4530 &0.4701 \\
& &100 &0.1 &0.5145 &0.5000 &0.4909 &0.4899 &0.4899 &0.4566 &0.4943 \\
& & &0.2 &0.5489 &0.5255 &0.5131 &0.5000 &0.4979 &0.5041 &0.5000 \\
& & &0.3 &0.5664 &0.5179 &0.4889 &0.5032 &0.5032 &0.5201 &0.5354 \\
0.9 &0.1 & 20 &0.1 &0.2738 &0.3396 &0.3586 &0.2975 &0.2908 &1.0000 &0.7512 \\
& & &0.2 &0.3386 &0.3634 &0.3889 &0.3069 &0.3139 &0.5158 &0.5529 \\
& & &0.3 &0.3815 &0.3529 &0.3411 &0.2572 &0.2546 &0.5763 &0.4110 \\
& & 50 &0.1 &0.4041 &0.4265 &0.4471 &0.4556 &0.4576 &1.0000 &0.5780 \\
& & &0.2 &0.4433 &0.4536 &0.4641 &0.4339 &0.4309 &0.5307 &0.4974 \\
& & &0.3 &0.4712 &0.4551 &0.4474 &0.3876 &0.3806 &0.5169 &0.4825 \\
& &100 &0.1 &0.4549 &0.4782 &0.4929 &0.4658 &0.4638 &0.5418 &0.4883 \\
& & &0.2 &0.4733 &0.4833 &0.4831 &0.4435 &0.4425 &0.5218 &0.4960 \\
& & &0.3 &0.4793 &0.4620 &0.4536 &0.4355 &0.4359 &0.4839 &0.4962 \\
&0.2 & 20 &0.1 &0.5271 &0.5234 &0.5235 &0.4862 &0.4836 &1.0000 &0.7331 \\
& & &0.2 &0.5217 &0.4655 &0.4433 &0.4250 &0.4319 &1.0000 &0.6052 \\
& & &0.3 &0.5325 &0.4548 &0.4249 &0.3779 &0.3652 &0.3757 &0.4572 \\
& & 50 &0.1 &0.4872 &0.4839 &0.4807 &0.4447 &0.4465 &0.4854 &0.5246 \\
& & &0.2 &0.5054 &0.4773 &0.4619 &0.4196 &0.4188 &0.4020 &0.4419 \\
& & &0.3 &0.5301 &0.4504 &0.4263 &0.4371 &0.4318 &0.5209 &0.4947 \\
& &100 &0.1 &0.4829 &0.4798 &0.4767 &0.4823 &0.4799 &0.4462 &0.4829 \\
& & &0.2 &0.5401 &0.5174 &0.5077 &0.4844 &0.4876 &0.4585 &0.4890 \\
& & &0.3 &0.5458 &0.5088 &0.4882 &0.4867 &0.4867 &0.4925 &0.5092 \\
&0.3 & 20 &0.1 &0.5102 &0.4827 &0.4709 &0.4579 &0.4569 &1.0000 &0.7051 \\
& & &0.2 &0.5493 &0.4858 &0.4685 &0.4556 &0.4461 &0.4949 &0.6437 \\
& & &0.3 &0.5799 &0.4094 &0.4500 &0.3883 &0.3808 &0.0000 &0.5165 \\
& & 50 &0.1 &0.4746 &0.4612 &0.4492 &0.4520 &0.4515 &0.4347 &0.4721 \\
& & &0.2 &0.5800 &0.5451 &0.5267 &0.5000 &0.4990 &0.4971 &0.5268 \\
& & &0.3 &0.5402 &0.4710 &0.4587 &0.4517 &0.4495 &0.5054 &0.4965 \\
& &100 &0.1 &0.5292 &0.5198 &0.5129 &0.4916 &0.4916 &0.4962 &0.4933 \\
& & &0.2 &0.5149 &0.4856 &0.4730 &0.4665 &0.4654 &0.5198 &0.4885 \\
& & &0.3 &0.5410 &0.5000 &0.4775 &0.4597 &0.4607 &0.4698 &0.4718 \\
 \bottomrule
\end{longtblr}
}

\begin{figure}[htpb]
  \centering
  \includegraphics[scale=0.36]{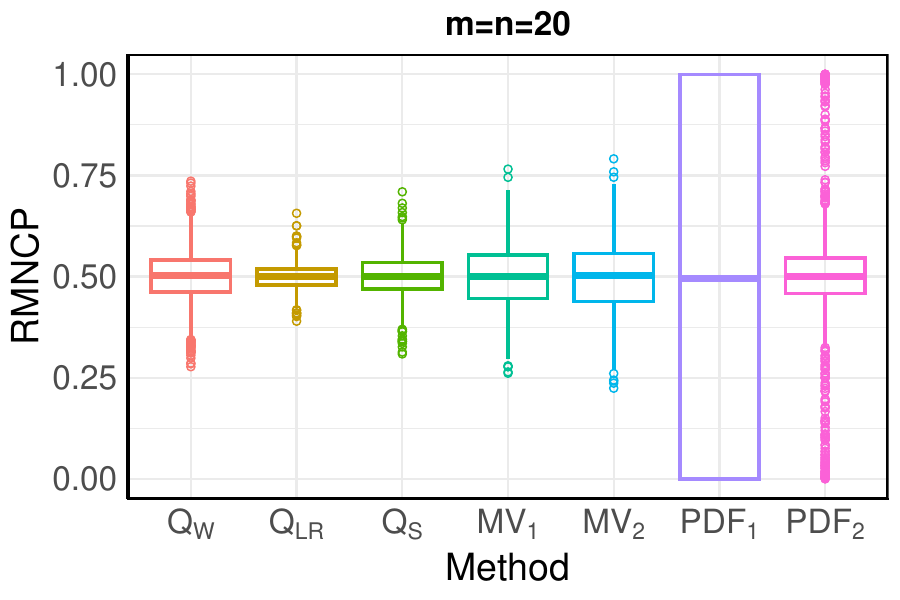}
  \includegraphics[scale=0.36]{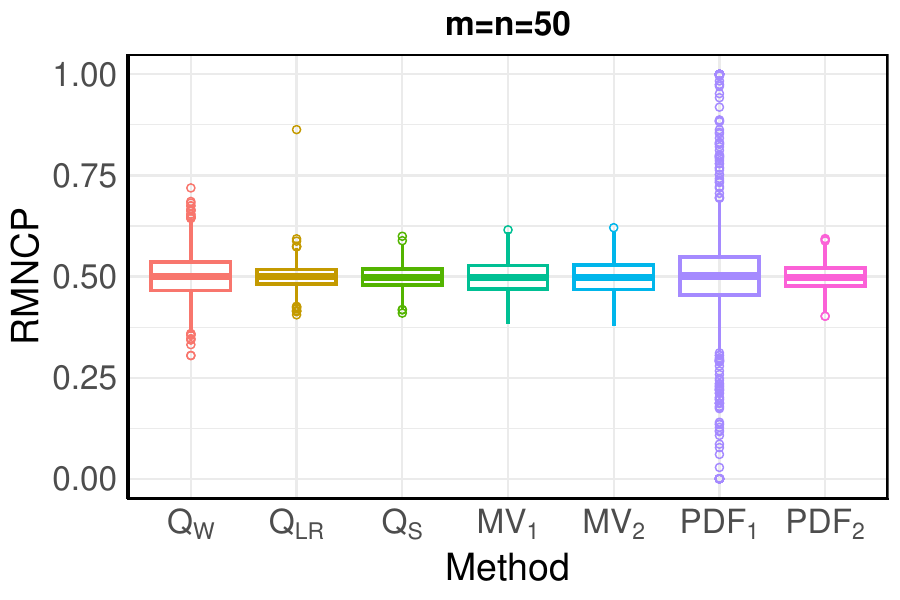}
  \includegraphics[scale=0.36]{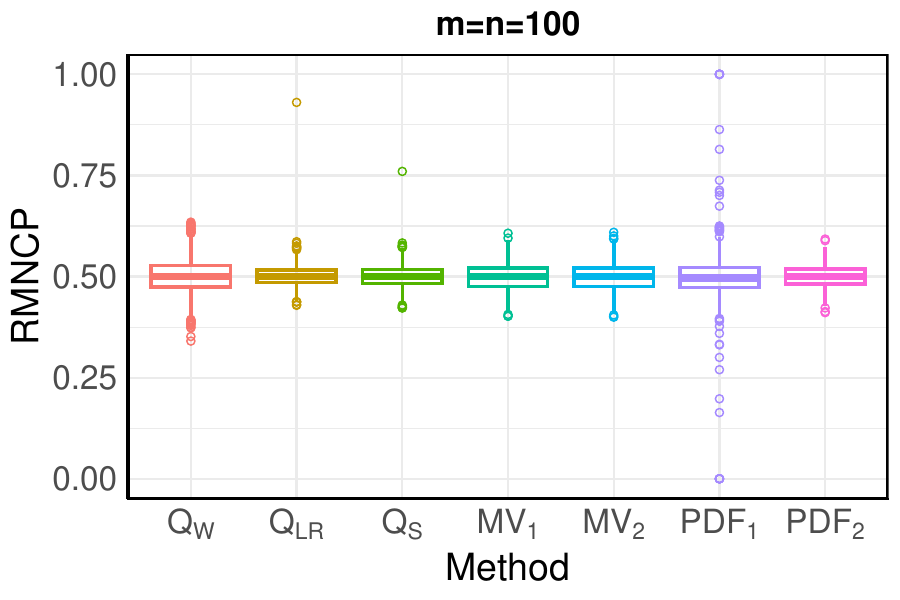}
  \caption{Box plots of the ratio of mesial non-coverage probability to the distal non-coverage probability (RMNCP) for the methods under consideration. The panels correspond to $m=n=20,50$ and $100$, respectively.}
  \label{fig:boxplot:rmncp}
\end{figure}

Figure~\ref{fig:boxplot:rmncp} displays box plots of the RMNCPs obtained from $1,000$ randomly generated admissible parameter configurations for $m=20,50$ and $100$. Across all method, the median RMNCP remains close to $0.5$, suggesting no systematic preference for either tail. However, when $m=n=20$, the distribution-based procedures exhibit substantially greater variability, with the interquartile range of $\mathrm{PDF}_1$ spanning nearly the entire interval from $0$ to $1$ and numerous extreme values observed for $\mathrm{PDF}_2$. As the sample size increases, the variability of the RMNCPs associated with the distribution-based methods decrease markedly. By $m=n=100$, the box plots of all methods are broadly similar and concentrated around $0.5$. 

Overall, all methods achieve satisfactory coverage performance, with departures from the nominal level occurring primarily under small sample settings and low baseline probabilities. The proposed distribution-based procedures generally produce narrower confidence intervals than the competing methods, particularly when the intra-subject correlation is low to moderate. The RMNCP results further suggest that the distribution-based approach is capable of accommodating asymmetry in the finite-sample distribution of the risk difference estimator, whereas the asymptotic and MOVER procedures tend to yield more balanced tail allocations due to their reliance on asymptotic normality. As the sample size increases, the differences among the competing methods become less pronounced, with all procedures exhibiting similar operating characteristics.

\subsection{Real-World Example}
\label{sec:results:real}
Two real examples are presented to illustrate the application of constructing confidence intervals for the risk difference. 

The first example involves a subset of 214 children who were admitted with acute otitis media with effusion (OME) and randomized into two treatment groups receiving cefaclor or amoxicillin, respectively~\cite{mandel1982duration}. Table~\ref{tab:OME} presents the number of cured ears among 173 children evaluated at 42 days.
\begin{table}[thpb]
    \centering
    \caption{Number of cured ears at 42 days in children treated with cefaclor and amoxicillin.}
    \label{tab:OME}
    \begin{tabular}{cccc}
    \toprule
         &\multicolumn{2}{c}{Treatment} & \\
         \cmidrule{2-3}
        \# of cured ears &Cefaclor &Amoxicillin &total \\
        \midrule
        0 &9 &7 &16 \\
        1 &7 &5 &12 \\
        2 &23 &13 &36 \\
        total &39 &25 &64 \\
        \midrule
        0 &20 &19 &39 \\
        1 &34 &36 &70 \\
        total &54 &55 &109 \\
        \bottomrule
    \end{tabular}
\end{table}

Goodness-of-fit tests proposed by Zhou and Ma~\cite{zhou2025goodness} indicate that Donner's model provides an adequate fit for the OME dataset, with all five recommended tests yielding $p$-values greater than $0.9$.
Table~\ref{tab:OME:stat} presents the estimated risk differences (RDs), 95\% CIs, and corresponding interval widths across different methods.
For the distribution-based approach, profile MLEs were used as plug-in values for the nuisance parameters because their true values are unknown in practice. The three asymptotic methods produce the point estimate of risk difference $\mathrm{RD}=-0.0119$, the MOVER gives $\mathrm{RD}=-0.0214$, and the distribution-based method yields $\mathrm{RD}=-0.0210$. Small differences in the point estimates arise because the asymptotic methods are based on the global MLE, whereas the MOVER-based and distribution-based approaches use the adjusted estimators $\tilde{p}_i$ in (\ref{eq:agresti-coull}) and $\tilde{\pi}_i$ in (\ref{eq:pis}), respectively. The resulting 95\% CIs are highly similar across methods and all contain zero. Consequently, we fail to reject the null hypothesis $H_0:~\delta=0$ and conclude that there is insufficient evidence of a difference in cure proportions between the cefaclor and amoxicillin treatment groups. The interval widths are comparable across methods, ranging from 0.2708 to 0.2766. The score-based CI yields the narrowest interval, whereas the two MOVER-based intervals are slightly wider. 
\begin{table}[thpb]
  \centering
  \caption{Example 1: Risk differences, 95\% CIs and the corresponding interval widths.}
  \label{tab:OME:stat}
  \begin{tabular}{lccc}
    \toprule
    Method &RD &95\% CI &Width \\
    \midrule
    $Q_W$ &-0.0119 &$\left[~-0.1481,0.1243~\right]$ &0.2725 \\
    $Q_{LR}$ &-0.0119 &$\left[~-0.1482,0.1235~\right]$ &0.2717 \\
    $Q_S$ &-0.0119 &$\left[~-0.1479,0.1229~\right]$ &0.2708 \\
    $\mathrm{MV}_1$ &-0.0214 &$\left[~-0.1578,0.1183~\right]$ &0.2762 \\
    $\mathrm{MV}_2$ &-0.0214 &$\left[~-0.1581,0.1185~\right]$ &0.2766 \\
    $\mathrm{PDF}_2$ &-0.0210 &$\left[~-0.1609,0.1141~\right]$ &0.2750 \\
    \bottomrule
  \end{tabular}
\end{table}

The second example concerns an observational study for sixty myopia patients receiving the so-called Orthokeratology (Ortho-k), a non-surgical vision correction method that uses specialized contact lenses worn overnight to temporarily reshape the cornea and correct myopia \cite{liang2024homogeneity}. There are two lens designs regarding Orth-k treatment method: (i) corneal refractive therapy (CRT) used by brand S; (ii) vision shaping treatment (VST) used by other brands. Myopia improvement is assessed by the axial length growth (ALG), where improvement is indicated if ALG is less than $0.3$ mm, and absent otherwise. The observations on the number of improved myopic eyes by lens designs are summarized in Table \ref{tab:myopia}. As can be seen, the dataset exhibits substantial sparsity, particularly in the unilateral portion where only six observations are available. This feature makes the dataset a useful example for assessing the robustness of the competing methods in a small sample setting. 
\begin{table}[thpb]
    \centering
    \caption{Number of improved myopic eyes with Ortho-k treatment using VST and CRT lens designs.}
    \label{tab:myopia}
    \begin{tabular}{cccc}
    \hline
         &\multicolumn{2}{c}{Lens Design} & \\
         \cline{2-3}
        \# of myopia improved eyes &VST &CRT &total \\
        \hline
        0 &20 &13 &33 \\
        1 &7 &2 &9 \\
        2 &10 &2 &12 \\
        total &37 &17 &54 \\
        \hline
        0 &3 &0 &3 \\
        1 &3 &0 &3 \\
        total &6 &0 &6 \\
        \hline
    \end{tabular}
\end{table}


The same goodness-of-fit tests were performed for the Ortho-k dataset under Donner's model. All five recommended methods yielded $p$-values greater than $0.9$, indicating that Donner's model provides an adequate fit. Table~\ref{tab:myopia:stat} presents the estimated risk differences (RDs), 95\% CIs, and corresponding interval widths obtained using the competing methods. Similar to the OME example, the three asymptotic methods yield identical point estimates, whereas the MOVER-based and distribution-based approaches produce slightly different estimates because they rely on adjusted estimators. The resulting 95\% CIs are highly comparable and all contain zero. Therefore, all methods lead to the same inferential conclusion, namely failure to reject the null hypothesis $H_0:~\delta=0$ at the 5\% significance level. There is insufficient evidence of a difference in the proportions of improved myopic eyes between the two lens designs. The interval widths are highly similar, ranging from 0.4109 to 0.4207. Notably, the Wald interval is the narrowest and only marginally includes zero, whereas the remaining methods yield slightly wider intervals with larger positive upper limits.
\begin{table}[thpb]
  \centering
  \caption{Example 2: Risk differences, 95\% CIs and the corresponding interval widths.}
  \label{tab:myopia:stat}
  \begin{tabular}{lccc}
    \toprule
    Method &RD &95\% CI &Width \\
    \midrule
    $Q_W$ &-0.2039 &$\left[~-0.4093,0.0015~\right]$ &0.4109 \\
    $Q_{LR}$ &-0.2039 &$\left[~-0.3921,0.0224~\right]$ &0.4146 \\
    $Q_S$ &-0.2039 &$\left[~-0.3859,0.0312~\right]$ &0.4171 \\
    $\mathrm{MV}_1$ &-0.1714 &$\left[~-0.3788,0.0343~\right]$ &0.4131 \\
    $\mathrm{MV}_2$ &-0.1714 &$\left[~-0.3822,0.0384~\right]$ &0.4207 \\
    $\mathrm{PDF}_2$ &-0.1985 &$\left[~-0.3766,0.0397~\right]$ &0.4163 \\
    \bottomrule
  \end{tabular}
\end{table}

To facilitate the application of the proposed and existing methods in practice, we provide a user-friendly online calculator available at~\href{https://www.buffalo.edu/~cxma/CI_RiskDiffRhoModelCombined.htm}{https://www.buffalo.edu/~cxma/CI\_RiskDiffRhoModelCombined.htm}. The calculator allows users to obtain confidence intervals using any of the six methods for either built-in example datasets derived from real-world applications and simulation studies or their own datasets. Users may also specify the desired confidence level $\brc{1-\alpha}$.

\section{Conclusions}
\label{sec:conclusions}
This paper investigated confidence interval construction for the risk difference between two proportions in combined unilateral and bilateral binary outcomes. Existing confidence interval procedures for this setting, including asymptotic and the MOVER-based methods, rely on asymptotic normality and may therefore be affected by finite-sample performance. 
To address this issue, we proposed a distribution-based confidence interval that utilizes the probability distribution of the risk difference estimator under Donner's model. The proposed approach was implemented through numerical inversion of the probability density function obtained from the characteristic function of the estimator. In addition, we adjusted the MOVER procedures by taking into account the intra-subject correlation effect which was not considered in the original version.

Extensive simulation studies were conducted to compare the proposed method with Wald-type-, likelihood ratio-, score- and MOVER-based confidence intervals. Across a broad range of parameter configurations, all methods exhibited satisfactory performance as the sample size increased. The proposed distribution-based interval generally achieved coverage probabilities close to the nominal level whiling maintaining interval widths comparable to those of existing methods.
While its coverage performance was not uniformly superior to that of the competing methods, the proposed approach was capable of capturing finite-sample distributional features, particularly skewness in certain parameter configurations, that were largely absent from the asymptotic approximations.
In small sample settings, the proposed approach was able to reflect skewness in the sampling distribution of the estimator in certain parameter configurations. This behavior was evident from the RMNCP analyses, where the distribution-based intervals produced non-central coverage patterns, whereas the competing methods often yielded nearly symmetric patterns centered around 0.5. 
The adjusted MOVER intervals generally exhibited performance comparable to that of the asymptotic methods.

Two real-world datasets from otolaryngologic and ophthalmologic studies were used to illustrate the application of the competing methods. In both examples, all methods produced similar confidence intervals and led to the same scientific conclusions. These analyses demonstrate the practical applicability of the proposed approach and suggest that it may serve as a useful alternative to existing procedures for inference on the risk difference.

As mentioned earlier, the proposed distribution-based approach can be applied under different statistical models in addition to Donner's model. A natural extension of this work is to explore the performance of the competing methods in alternative dependence models for bilateral outcomes. Furthermore, one may extend the framework to consider other effect measures, such as the risk ratio and odds ratio. 
We also note that the distribution-based approach may exhibit slightly lower coverage accuracy in certain small sample scenarios despite providing a more realistic representation of skewness in the sampling distribution. This phenomenon may be attributable to numerical evaluation of the probability density function when the estimator possesses a highly discrete distribution in small samples. 
Additional research on computational improvements and higher order refinements may further enhance the applicability of distribution-based inference for correlated binary data.

\appendix
\section{Characteristic Function for Risk Difference Estimator}\label{app:CF}
Conditioning on the total counts ($m_{+i},~n_{+i}$) in each group, the characteristic function~(\ref{eq:cf}) is derived as follows.
\begin{equation}
  \begin{aligned}
    \varphi_{\tilde{\delta}}\brc{t}
    &=\mathbb{E}\brc{e^{i\tilde{\delta}}} \\
    &=\mathbb{E}\brc{e^{i\tilde{\pi}_2}}\mathbb{E}\brc{e^{-i\tilde{\pi}_1}} \\
    &=\mathbb{E}\brc{e^{i\frac{tm_{12}+2tm_{22}+tn_{12}}{2m_{+2}+n_{+2}}}}\mathbb{E}\brc{e^{-i\frac{tm_{11}+2tm_{21}+tn_{11}}{2m_{+1}+n_{+1}}}} \\
    &=\mathbb{E}\brc{e^{i\bt_2\cdot\bm_2}}\mathbb{E}\brc{e^{i\bt_2'\cdot\bn_{2}}}\mathbb{E}\brc{e^{i\bt_1\cdot\bm_1}}\mathbb{E}\brc{e^{i\bt_1'\cdot\bn_{1}}}, 
  \end{aligned}
\end{equation}
where
\begin{align*}
  &\bm_k=\brc{m_{0k},m_{1k},m_{2k}}, \quad \bn_k=\brc{n_{0k},n_{1k}}, \\
  &\bt_k=\frac{s_k}{D_k}\brc{0,t,2t}, \quad \bt_k'=\frac{s_k}{D_k}\brc{0,t}, \\
  &s_k=\brc{-1}^k, \quad D_k=2m_{+k}+n_{+k}, 
\end{align*}
for $k=1,2$.

Using the characteristic function for multinomial distribution that
\begin{equation*}
  \varphi_{\bm}\brc{\bt}=\mathbb{E}\brc{e^{i\bt\cdot\bm}}=\brc{\sum_{j=0}^kp_je^{it_k}}^{m_+}, 
\end{equation*}
for $\bm=\brc{m_0,\ldots,m_l}\sim multinomial\brc{m_+,p_0,\ldots,p_l}$ and $\bt=\brc{t_0,\ldots,t_l}$, we have
\begin{equation}
  \begin{aligned}
    \mathbb{E}\brc{e^{i\bt_k\cdot\bm_k}}&=\brc{p_{0k}+p_{1k}e^{s_k\frac{it}{D_k}}+p_{2k}e^{s_k\frac{i2t}{D_k}}}^{m_{+k}}, \\
    \mathbb{E}\brc{e^{i\bt_k'\cdot\bn_k}}&=\brc{1-\pi_k+\pi_ke^{s_k\frac{it}{D_k}}}^{n_{+k}},
    \quad
    k=1,2. 
  \end{aligned}
\end{equation}

Thus,
\begin{equation}
  \varphi_{\tilde{\delta}}\brc{t}=\prod_{k=1}^2\brc{p_{0k}+p_{1k}e^{s_k\frac{it}{D_k}}+p_{2k}e^{s_k\frac{i2t}{D_k}}}^{m_{+k}}\brc{1-\pi_k+\pi_ke^{s_k\frac{it}{D_k}}}^{n_{+k}}. 
\end{equation}

Note that $\brc{\bm_i,\bn_i}\perp\brc{\bm_j,\bn_j}$ ($i\ne j$) and $\bm_i\perp\bn_i$ are implicitly assumed, i.e., outcomes from different groups and/or subjects are independent.

\vspace{2em}
\declare{Author contributions}{The authors confirm contribution to the paper as follows: study conception and design: Zhou J, Ma C-X; analysis and interpretation of results: Zhou J, Ma C-X; draft manuscript preparation: Zhou J. All authors reviewed the results and approved the final version of the manuscript.}

\declare{Conflict of interest}{The authors declare that they have no conflict of interest.}

\declare{Funding}{This research received no external funding.}

\declare{Data availability}{The datasets analyzed in this study are publicly available in references~\cite{mandel1982duration,liang2024homogeneity}. To facilitate practical application of the proposed and existing methods, an online calculator for computing confidence intervals is available at~\href{https://www.buffalo.edu/~cxma/CI_RiskDiffRhoModelCombined.htm}{https://www.buffalo.edu/~cxma/CI\_RiskDiffRhoModelCombined.htm}.}

\bibliographystyle{unsrt}
\bibliography{correlateddata}

\end{document}